\documentclass{jfm}
\usepackage{graphicx}
\usepackage{xcolor}
\usepackage{epstopdf, epsfig}

\usepackage{hyperref}

\usepackage{color}  
\definecolor{darkGreen}{rgb}{0,0.45,0}
\definecolor{darkBlue}{rgb}{0,0,0.9}
\definecolor{darkRed}{rgb}{0.76, 0.13, 0.28}
\definecolor{darkPurple}{rgb}{.5, 0, 1}
\definecolor{darkOrange}{rgb}{0.8, 0.33, 0.0}
\hypersetup{
	colorlinks=true, 
	linktoc=all,    
	linkcolor=darkBlue, 
	citecolor=darkBlue,
	urlcolor = darkRed,
}
\usepackage{makecell}
\usepackage{hhline}
\usepackage{array}
\usepackage{tabulary}
\usepackage{multirow}
\newcolumntype{K}[1]{>{\centering\arraybackslash}m{#1}}

\newcommand\St{\mbox{\textit{St}}}  

\usepackage{mathtools} 

\def\Xint#1{\mathchoice
	{\XXint\displaystyle\textstyle{#1}}%
	{\XXint\textstyle\scriptstyle{#1}}%
	{\XXint\scriptstyle\scriptscriptstyle{#1}}%
	{\XXint\scriptscriptstyle\scriptscriptstyle{#1}}%
	\!\int}

 \def\XXint#1#2#3{{\setbox0=\hbox{$#1{#2#3}{\int}$}
\vcenter{\hbox{$#2#3$}}\kern-.5\wd0}}

\def\XXiint#1#2#3{{\setbox0=\hbox{$#1{#2{\mathrm{#3\!\! #3}}}{\iint}$}
\vcenter{\hbox{$#2 {\mathrm{#3\!\! #3}}$}}\kern-.5\wd0}}

\usepackage{subcaption}
\usepackage{mathtools}
\usepackage{enumerate}
\usepackage{wasysym}
\usepackage{stmaryrd}

\newcommand*{\iin}{\textup{in}}
\newcommand*{\oout}{\textup{out}}

\newcommand{\Si}{\mathrm{Si}}
\newcommand{\Ci}{\mathrm{Ci}}
\renewcommand{\d}{{\mathrm{d}}}
\usepackage{float}

\shorttitle{Interactions of tandem flapping wings}
\shortauthor{F. Fang, C. Mavroyiakoumou, L. Ristroph, and M. J. Shelley}

\title{Flow interactions and forward flight dynamics of tandem flapping wings}

\author{Fang Fang\aff{1},
  Christiana Mavroyiakoumou\aff{1} \corresp{\email{cm4291@nyu.edu}},
  Leif Ristroph\aff{1}
 \and Michael J. Shelley\aff{1,2}}

\affiliation{\aff{1}Courant Institute of Mathematical Sciences, 
New York University, 251 Mercer Street, New York, NY 10012, USA
\aff{2}Center for Computational Biology, Flatiron Institute,
162 Fifth Ave, New York, NY 10010, USA}

\begin{document}

\maketitle

\begin{abstract}
We examine theoretically the flow interactions and forward flight dynamics of tandem or in-line flapping wings. Two wings are driven vertically with prescribed heaving-and-plunging motions, and the horizontal propulsion speeds and positions are dynamically selected through aero- or hydro-dynamic interactions. Our simulations employ an improved vortex sheet method to solve for the locomotion of the pair within the collective flow field, and we identify `schooling states' in which the wings travel together with nearly constant separation. Multiple terminal configurations are achieved by varying the initial conditions, and the emergent separations are approximately integer multiples of the wavelength traced out by each wing. We explain the stability of these states by perturbing the follower and mapping out an effective potential for its position in the leader's wake. Each equilibrium position is stabilized since smaller separations are associated with in-phase follower-wake motions that constructively reinforce the flow but lead to decreased thrust on the follower; larger separations are associated with antagonistic follower-wake motions, increased thrust, and a weakened collective wake. The equilibria and their stability are also corroborated by a linearized theory for the motion of the leader, the wake it produces, and its effect on the follower. We also consider a weakly-flapping follower driven with lower heaving amplitude than the leader. We identify `keep-up' conditions for which the wings may still `school' together despite their dissimilar kinematics, with the `freeloading' follower passively assuming a favorable position within the wake that permits it to travel significantly faster than it would in isolation.
\end{abstract}

\begin{keywords}
Authors should not enter keywords on the manuscript, as these must be chosen by the author during the online submission process and will then be added during the typesetting process (see http://journals.cambridge.org/data/\linebreak[3]relatedlink/jfm-\linebreak[3]keywords.pdf for the full list).
\end{keywords}

\section{Introduction}

Flow interactions are thought to play an important role in the locomotion of highly ordered animal groups, such as schools of fish and flight formations of birds. A popular belief is that the group is able to reduce on the energetic cost needed to move through the fluid by taking advantage of the collectively-generated flows. For example, individual birds in a V-formation are thought to exploit the upward flow or upwash generated by neighboring birds to generate the required lift with reduced effort \citep{Lissaman1970Formationflightbirds,
Weimerskirch2001Energysavingin,
Portugal2014Upwashexploitationand}. For fish schools, an increasing number of studies show that individuals in a group gain energetic benefits compared to swimming alone \citep{Belyayev1969Hydrodynamichypothesisschool,
Zuyev1970experimentalstudyswimming,
Svendsen2003Intraschoolpositional,
Killen2011Aerobiccapacityinfluences}.
An early theory argues that each fish extracts energy from the vortical flows created by others, and predicts a particularly favorable arrangement is a diamond-shaped lattice \citep{Weihs1973Hydrodynamicsfishschooling,
Weihs1975Somehydrodynamicalaspects}. Although this optimal configuration has not been observed in field and laboratory studies \citep{Partridge1979evidenceagainsthydrodynamic}, efficiency improvement due to vortex-swimmer interactions has been seen in  modeling and simulation studies \citep{Hemelrijk2015increasedefficiencyfish,
Daghooghi2015hydrodynamicadvantagessynchronized}.

Studies have made progress by considering the simplified problem of a single body swimming or flying within ambient vortices. Experiments have shown that trout swimming in a von K\'arm\'an wake behind a bluff body tend to slalom between vortex cores, and measurements showing reduced muscle activity suggest that flow effects are being exploited \citep{Liao2003Fishexploitingvortices,Liao2007reviewfishswimming}. Theoretical modeling has examined a flexible filament swimming in a vortex street, where the thrust or efficiency can be maximized with respect to the flapping kinematics \citep{Alben2009swimmingflexiblebody,Alben2010Passiveandactive}. Other studies focus on a pair of flapping and interacting bodies, for example, a pair of passively flapping filaments \citep{Ristroph2008Anomaloushydrodynamicdrafting,
Alben2009Wakemediatedsynchronization,
Alben2012Flappingpropulsionusing}, as well as actively flapping wings or foils placed side-by-side \citep{Dewey2014Propulsiveperformanceunsteady} or in tandem \citep{Akhtar2007Hydrodynamicsbiologicallyinspired,
Boschitsch2014Propulsiveperformanceunsteady}. Interactions of freely swimming and actively flapping bodies were studied in CFD simulations \citep{Zhu2014Flowmediatedinteractions}. A pair of actively flapping and flexible filaments swimming freely in tandem revealed the formation of stable configurations of reduced energetic input. Although \citet{Zhu2014Flowmediatedinteractions} used a somewhat low Reynolds number ($\Rey=200$), similar stable configurations of tandem flapping hydrofoils were observed in 
experiments at the higher Reynolds numbers ($\Rey=10^4 \sim 10^5$) typical of schools and flocks \citep{Becker2015Hydrodynamicschoolingflapping, Ramananarivo2016Flowinteractionslead}.
 
By using freely-swimming and flapping hydrofoils in controlled experiments, \citet{Becker2015Hydrodynamicschoolingflapping,Ramananarivo2016Flowinteractionslead,newbolt2019flow} and \citet{newbolt2024flow} sought to understand the flow interactions relevant to high-$Re$ collective locomotion. These studies build on previous ones that have characterized the locomotion of a single heaving-and-plunging foil \citep{Vandenberghe2004Symmetrybreakingleads, Alben2005Coherentlocomotionas, 
Vandenberghe2006unidirectionalflightfree}. Such a locomotor leaves a stereotypical thrust wake consisting of an array of staggered and counter-rotating vortices, similar to that seen behind swimming fish \citep{Mueller1997Fishfootprints}. 
\citet{Becker2015Hydrodynamicschoolingflapping} studied the hydrodynamic interactions and swimming dynamics of an in-line array of flapping foils held at fixed separation distances. 
Constructive and destructive wing-wake interactions were found to coexist and correspond to fast and slow swimming modes of the group, respectively. Introducing the so-called schooling number $S$, defined as the inter-neighbor separation distance normalized by the wavelength of the swimming trajectory, it was observed that the array assumed $S$ values only between integer to half integer values, which may be associated with stable wing-wake interactions. 
This was investigated theoretically by \citet{oza2019lattices} using a discrete-time dynamical system, 
which showed good agreement with the
experimental data, in particular predicting the bistability of schooling states.
The importance of hydrodynamic stability in collective locomotion, heretofore rarely discussed in the literature, was further investigated by \citet{Ramananarivo2016Flowinteractionslead} for the case of tandem flapping foils. In this experimental realization of the simplest `school of two', the wings are freely swimming and freely spacing. The emergent stable configurations involve the pair traveling together with constant separation and with integer schooling numbers, i.e. the wings maintain a separation distance that is an integer multiple of the motion wavelength.
\citet{Ramananarivo2016Flowinteractionslead} also measured directly the hydrodynamic restoring forces on the follower and for the first time determined the hydrodynamic potential felt by a locomotor interacting with the wake of a leader. \citet{newbolt2019flow} 
investigated the flow interactions between uncoordinated flapping foils with different kinematics, showing that these interactions can spontaneously lead to group cohesion.
A recent experimental study involving larger collectives (of up to five foils) demonstrated that chains of increasing size become unstable due to flow-induced instabilities termed flonons~\citep{newbolt2024flow}. In this phenomenon, the horizontal positions of downstream foils begin to oscillate with progressively larger amplitudes down the group, potentially leading to collisions between the trailing members. This instability was subsequently found numerically using a vortex sheet model in \citet{nitsche2024stability}.

\begin{figure}
\centering
\includegraphics[width=\textwidth]{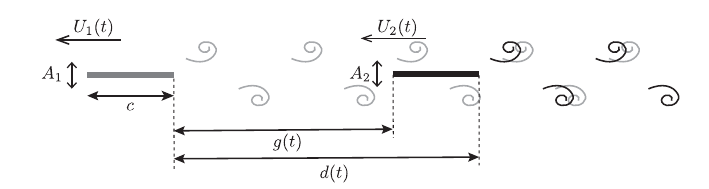}\caption{Interaction of a pair of flapping wings in tandem. Two wings, modeled as slender rigid plates, heave and plunge with prescribed vertical motions and are individually free to translate in the horizontal direction. Without loss of generality, we assume the wings swim or fly from right to left. Here $A_k$ is the prescribed heaving amplitude, $U_k$ is the emergent swimming speed (with $k=1,2$), $c$ is the chord length of each of the rigid plates, $g$ is the ``tail-head" separation distance, and $d$ is the separation distance between the wing center points (or any equivalent points on the two wings).}
\label{fig:set_up}
\end{figure}

Inspired by these previous studies, we investigate theoretically the interactions and collective dynamics of a tandem pair of free flapping wings swimming or flying through a fluid. In simulations and a model, we consider two identical and infinitesimally-thin plates that are driven up and down vertically (i.e. heaving and plunging) in a 2D inviscid and incompressible flow (see figure~\ref{fig:set_up}). A follower wing is placed directly downstream of a leader, and the two have identical flapping frequency and temporal phase but perhaps different amplitudes. The wings are individually self-propelled and swim or fly in the horizontal direction, and the two are coupled only through their collective flow field. 

In the first half of this paper, we discuss numerical simulations that use a two-dimensional (2D) vortex sheet method to study fluid-structure interactions at high~$\Rey$. We follow the scheme described in \citet{Fang2016computationalmodelflight} that incorporates the fast multipole method (FMM) to efficiently simulate the complex vortical flow fields generated by a dynamic body, and its interaction with these flows. The quadrature rules are improved by applying singularity subtractions and piece-wise analytic evaluations to prevent vortex sheet penetration at the wing boundaries. The Blasius skin-friction calculation \citep{Schlichting1968Boundarylayertheory} is coupled with this scheme as a model of viscous drag on each wing, permitting the study of steady states set by force balance. Our simulations confirm previous numerical and experimental results \citep{Zhu2014Flowmediatedinteractions, Ramananarivo2016Flowinteractionslead,newbolt2019flow,heydari2021school} of multiple stable configurations, called `schooling states' in what follows. We also map out an effective hydrodynamic potential on the follower wing using two strategies, and find the hydrodynamic force on the follower is approximately a sinusoidal function of the schooling number $S$, defined as the tail-to-head wing separation normalized by the locomotion wavelength. The hydrodynamic potential consists of multiple wells corresponding to the multiple stable states. Investigation of the emergent flow fields shows that when the follower's flapping motion is in-phase with the oscillatory flow left by the leader, the collective wake is constructively reinforced but the follower produces low thrust. In contrast, for out-of-phase motions with the oncoming flow, the follower produces larger thrust as the collective wake is weakened. We show how these effects can be exploited by a follower with low flapping amplitude swimming behind a fast-flapping leader. Surprisingly, this `freeloading' follower is still able to keep up with the leader, as it passively relocates to a favorable position in the leader's wake. A significant improvement in swimming efficiency is observed in this situation, with the follower even harvesting energy from the wake of the leader.

In the second half of this paper, we calculate analytically the stable states of the two-wing interaction and the hydrodynamic potential on the follower wing using a linearized theory that assumes small flapping amplitude and large wavelength of motion. 
The linearized Euler equation with a no-penetration boundary condition imposed on a plate was analytically solved by \citet{Wu1961Swimmingwavingplate}. We combine Wu's solution with a skin-friction model to determine the free swimming speed of an isolated flapping wing. Motivated by the simulation results, we use the calculations of the isolated wing and its generated wake as approximations to the leader wing and its wake. We then adapt a second calculation of Wu for the thrust on a flapping wing in a wavy stream \citep{Wu1971Extractionflowenergy,Wu1975Extractionflowenergy} to the interaction of the follower wing with the wavy wake generated by the leader. We provide an analytic expression of the effective hydrodynamic potential on the follower wing, and stable states and the `keep-up' condition for a low-amplitude follower are also determined.

\section{Simulation method} \label{sec: simulation method}
Efficient simulations of high Reynolds number flows are challenging due to the associated complex vortex dynamics \citep{Saffman1993VortexDynamics}. In our study, we use an inviscid 2D vortex sheet model to capture both the vorticity distribution along the wings (called the bound sheet, as defined later) and the free vortex wake (the free sheet), as well as the unsteady shedding of vorticity from the wing trailing edge. Within this 2D model, a {\it vortex sheet} is a 1D boundary across which the fluid normal velocity is continuous while the tangential velocity is not \citep{Rosenhead1931formationvorticesfrom, Saffman1993VortexDynamics}. A stabilized 2D vortex sheet model was first proposed by \citet{Krasny1986Desingularizationperiodicvortex} using a kernel regularization technique. Combined with the unsteady Kutta condition, this method is widely used for fluid-structure interaction problems at high \Rey, for example in simulations of a jet flow exiting a tube \citep{Nitsche1994numericalstudyvortex}, free falling plates
 \citep{Jones2003separatedflowinviscid, Jones2005Fallingcards}, passively flapping filaments \citep{Alben2008Flappingstatesflag,Alben2009Simulatingdynamicsflexible}, and free flying jellyfish-like machines \citep{Fang2016computationalmodelflight}. 
In this work, we build on the vortex sheet formulation for wing-pair interactions and free locomotion described by \citet{Fang2016computationalmodelflight}, and we implement a robust numerical algorithm using the fast multipole method (FMM). To account for viscous effects and drag, we also include a Blasius skin-friction model \citep{Schlichting1968Boundarylayertheory}.  For each individual wing, the balance of the skin-friction drag and computed thrust from the vortex sheet determines the dynamics of the wing in the horizontal direction. We also improve the quadrature rules in the vortex sheet method, by applying singularity subtractions on the body and piece-wise analytic evaluations on the wake, so as to prevent vortex sheet penetration at the wing boundaries.

\subsection{Modeling}\label{subsec:simulation model}

\subsubsection{Wing model and fluid model} 
As shown in figure~\ref{fig:set_up}, the pair of wings in tandem are each  modeled as rigid plates of the same chord length $c$. The up-and-down or heaving-and-plunging motions are prescribed by sinusoidal functions of the same frequency $f$ and (perhaps different) amplitudes $A_k$,
\begin{eqnarray}
x_{k}(s,t) & = & X_k(t)+s, \; s\in[-c/2,c/2],\\
y_{k}(t) & = & A_k\cos(2\pi ft),\label{Eq: heaving}
\end{eqnarray}
where $k=1,2$ denotes the leader and follower wings, respectively, and $\mathbf{x}_k=(x_k,y_k)$ denotes their locations. Here $y_k(t)$ is the same throughout the wing's chord length (i.e. for any $s\in[-c/2,c/2]$ of each wing), and so it is a function of time only. The arclength parameter $s$ denotes the signed distance from the wing center ${X_k(t)= x_k(0,t)}$. The leading edge of the wings (left end of the wing in figure~\ref{fig:set_up}) corresponds to $s=-c/2$, and the trailing edge (right end) corresponds to $s=c/2$.

The wings are individually free to swim or fly, that is, to move in the horizontal direction due to hydrodynamic interactions. Assuming a wing mass of $m$, the horizontal dynamics of each individual satisfies,
\begin{equation}\label{eq:momentum}
m\ddot{X}_k = F_k = T_k+D_k,\quad k=1,2,
\end{equation}
where $F_k$ is the hydrodynamic force on the wing, composed of the hydrodynamic thrust~$T_k$ and drag $D_k$. Here, the propulsive thrust is negatively signed and the resistive drag is positively signed, given the convention that the wings swim or fly from right to left.

The surrounding inviscid fluid flow is described by the 2D Euler equations,
\begin{equation}\label{Eq:Euler}
\rho_{f}\frac{D\mathbf{u}}{Dt}=-\nabla p,
\end{equation}
where $\mathbf{u}=(u_x,u_y)$ is the fluid velocity, $p$ is the pressure and $\rho_{f}$ is the fluid density. 
We impose a no-penetration boundary condition for the flow velocity $\mathbf{u}(\mathbf{x}(s))$ on each wing, with the vertical component of the velocity being continuous and matching the wing flapping speed. That is, the vertical component $u_{y,\pm}$ above and below the wing are equal to the vertical component of the wing velocity:
\begin{eqnarray}\label{eq:bc}
u_{y,+}(\mathbf{x}_k(s,t))=u_{y,-}(\mathbf{x}_k(s,t))=\dot{y}_k(t),\quad k=1,2.
\end{eqnarray}
In the tangential direction at the wing, i.e. the $x$-direction, flow is allowed to slip freely along the surface. This horizontal component $u_{x,\pm}(\mathbf{x}_k(s,t))$ of the flow velocity can be viewed as a model for the flow just outside of the boundary layer, which approaches zero thickness in the limit of infinite Reynolds number. We later discuss a skin-friction model that estimates the drag due to shear in the viscous boundary layer.

Integration of the fluid pressure over a closed contour around each wing surface $C_k^w$ provides the thrust, i.e. the hydrodynamic force in the horizontal or tangential direction $\hat{\mathbf{s}}=(1,0)$.
This integral can be separated into parts associated with an infinitesimal circle (denoted by $r_k^{le}$) around the wing's leading edge, the differential or jump in pressure $[p]$ across the length of the wing, and an infinitesimal circle $r_k^{te}$ at the trailing edge:
\begin{equation}\label{eq:pressure_integral}
\int_{C_k^w}p\hat{\mathbf{n}}\,\d s \cdot \hat{\mathbf{s}}=\left ( \int_{r_k^{le}}p\hat{\mathbf{n}}\,\d s
+\int_{-1}^1[p]\hat{\mathbf{n}}\,\d s 
+ \int_{r_k^{te}}p\hat{\mathbf{n}}\,\d s \right) \cdot \hat{\mathbf{s}},
\quad \hat{\mathbf{s}}=(1,0). 
\end{equation}
In our model, we allow continuous shedding of vorticity at the trailing edge while keeping a flow singularity at the leading edge. The absence of leading edge shedding is intended as a model applicable to the case of small-amplitude flapping and long wavelengths of motion, in which case the small angles of attack are associated with weak leading edge separation. The singularity at the leading edge yields a leading-edge suction and thrust in the tangential direction to the wing \citep{Saffman1993VortexDynamics}. Note that the integral of the pressure jump $[p]$ along the wing is always normal to the wing and does not contribute to the thrust. The pressure integral around the wing trailing edge is also zero due to the imposition of the unsteady Kutta condition. Therefore, for a heaving-and-plunging wing that swims in a direction parallel to the body axis, the thrust results purely from leading-edge suction. Thus, Eq.~\eqref{eq:pressure_integral} simplifies to:
\begin{equation}\label{eq:Tk suction}
T_k =\int_{C_k^w}p\hat{\mathbf{n}}\,\d s \cdot \hat{\mathbf{s}}
=\int_{r_k^{le}}p\hat{\mathbf{n}}\,\d s \cdot \hat{\mathbf{s}},
\quad \hat{\mathbf{s}}=(1,0). 
\end{equation}

The hydrodynamic drag in Eq.~(\ref{eq:momentum}) results from skin friction along the upper and lower wing surfaces, which is modeled using Blasius laminar boundary layer theory \citep[][Chap.~VII]{Schlichting1968Boundarylayertheory}. 
For one side of a plate in a stream of uniform far-field speed $U_f$, this approximation gives \begin{equation}\label{Eq:Df}
D_f=0.664\sqrt{\rho_{f}\mu c}\,| U_f|^{3/2},
\end{equation}
where $\mu$ is the dynamic viscosity of the fluid. In our model, we replace the free stream velocity $U_f$ with the tangential flow speed averaged along the wing surface and relative to the swimming speed: 
\begin{equation}\label{Eq:U_kpm}
\overline{U}_{k,\pm}(t) = \frac{1}{c}\int_{-c/2}^{c/2}u_{x,\pm}(\mathbf{x}_k(s,t))\,\d s-U_k(t),
\end{equation} 
where $U_k(t)=\dot{X}_{k}(t)$ is the horizontal swimming speed of the wing, which is assumed to move in the negative $x$-direction (figure~\ref{fig:set_up}). Summing the contributions from the upper and lower surfaces of the wing, the total drag is
\begin{equation}\label{Eq:Dk}
D_k=0.664\sqrt{\rho_{f}\mu c}\,\left(|\overline{U}_{k,+}|^{3/2}+|\overline{U}_{k,-}|^{3/2}\right).
\end{equation}
This form of the drag force was also implemented by \citet{heydari2021school}.

We non-dimensionalize the system with the characteristic length $\tilde{L}=c/2,$ time ${\tilde{T}=1/f}$, velocity $\tilde{U}=cf/2$
and pressure $P=\rho_f\tilde{U}^2=\rho_{f}c^{2}f^{2}/4.$ The dimensionless equations of motion are then
\begin{eqnarray}\label{Eq:MX_k}
M\ddot{X}_k & = & T_k(t)+D_k(t), \\
y_{k}(s,t) & = & \tilde{A}_k\cos(2\pi t),\ s\in[-1,1],
\end{eqnarray}
where
\begin{equation}\label{Eq:TkDk}
T_k = \int_{C_k^w}p\hat{\mathbf{n}}\,\d s\cdot \hat{\mathbf{s}}\quad 
\text{and}\quad  D_k = C_f\left(\overline{U}_{k,+}^{3/2}+\overline{U}_{k,-}^{3/2}\right),\; \quad k=1,2.
\end{equation}

There are three dimensionless parameters that appear in the system: the wing-fluid mass ratio $M=4m/(\rho_fc^2)$, the ratio of the heaving amplitude to wing length ${\tilde{A}_k = 2A_k/c}$, and the skin-friction drag coefficient 
\begin{equation}\label{Eq:Cf}
C_f = 1.878\sqrt{\nu/(c^2f)},
\end{equation}
where $\nu=\mu/\rho_f$ is the kinematic viscosity. This drag coefficient is related to the so-called frequency Reynolds number $Re_{fr}=\tilde{L}\tilde{U}/\nu=c^2f/4\nu$ used in the literature of flapping-wing locomotion \citep{Alben2005Coherentlocomotionas}: $C_f=0.939Re_{fr}^{-1/2}$. As will be shown below, employing $C_{f}\sim 0.02$ yields swimming speeds comparable to those observed in the experiments of \citet{Ramananarivo2016Flowinteractionslead} for appropriately matched kinematics, wing chord length, and fluid parameters. The same value of $C_f$ is estimated by \citet{heydari2021school}. In our study, we set the mass ratio to $M=2$ while exploring the effects of varying $C_{f}=0.01$ to $0.1$ and $\tilde{A}_{k}=0.1$ to $1$.

\subsubsection{Vortex sheet shedding and wake model} \label{subsubsec:VortexSheetModel}
In the vortex sheet model, a flapping wing is a {\it bound vortex sheet} on which the no-penetration boundary condition Eq.~(\ref{eq:bc}) is imposed. The bound sheet encompasses the infinitesimally-thin wing and its boundary layers that vanish in thickness at higher Reynolds numbers. At the trailing edge of each wing, a {\it free vortex sheet} is continuously shed into the fluid \citep{Nitsche1994numericalstudyvortex}. In the free sheet models the shed shear layer forms downstream eddies. In the limit of zero viscosity, the shed vorticity concentrates onto an infinitesimally-thin layer that can be modeled as a 1D vortex sheet that does not dissipate in time. The vortex shedding rate at the trailing edge is determined by the unsteady Kutta condition \citep{Jones2003separatedflowinviscid}, which provides the direction of vortex shedding as well as the amount of circulation transmitted from the wing (bound  sheet) to the wake (free  sheet). 

In our 2D vortex sheet method, we keep track of the sheet position $\zeta_k(s,t)$ as well as the true vortex sheet strength $\bar{\gamma}_k(s,t)$, defined as the tangential fluid velocity jump across the vortex sheet $\bar{\gamma}_k(s,t)=[\mathbf{u}]\hat{s}$ \citep{Shelley1992studysingularityformation}. Here $k=1,2$ denotes the two wings, and $s\in[-1,s^k_{\max}]$ is an arc-length parameterization of the vortex sheet. The true vortex sheet strength $\bar{\gamma}_k(s,t)=\gamma_k(\alpha)/s_{\alpha}(\alpha,t)$ is related to the (unnormalized) vortex strength $\gamma_k(\alpha)$ which is a material variable, where $\alpha$ is a Lagrangian parameterization of the sheet. Using complex variables, the vortex sheet position $\zeta_k(s,t)$ and strength $\bar{\gamma}_k(s,t)$ are related through the Biot-Savart law for the mean velocity of the sheet $w_k(s,t)$,
\begin{equation}\label{eq:w_kst}
\overline{w_k(s,t)}  =  \frac{1}{2\pi i}\Xint-_{-1}^{s_{\max}^k}\frac{\bar{\gamma}_{k}(s',t)\,\d s'}{\zeta_{k}(s,t)-\zeta_{k}(s',t)}
+\sum_{j\neq k}\frac{1}{2\pi i}\int_{-1}^{s_{\max}^j}\frac{\bar{\gamma}_{j}(s',t)\,\d s'}{\zeta_{k}(s,t)-\zeta_{j}(s',t)}, 
\end{equation}
where the bar denotes the complex conjugate, and $\Xint-$ is the Cauchy principal value. Since the tangential component of fluid velocity is discontinuous at the vortex sheet, the mean velocity $w_k(s,t)$ is an average of the velocities on the two sides of the sheet boundary $\zeta_k(s,t)$.

The description of the free sheet boundary $s\in[1,s^k_{\max}]$ dynamics is simplified if $w_{k}(s,t)$ is chosen as a velocity frame, in which case we arrive at the Birkhoff--Rott equation \citep{Shelley1992studysingularityformation},
\begin{equation}\label{eq:freesheet}
D_t\zeta_{k}(s,t)	=w_{k}(s,t),\quad 1\le s\le s_{\max}^k,
\end{equation}
where $D_t$ denotes the material derivative, and $w_{k}(s,t)$ is given by Eq.~(\ref{eq:w_kst}). Following the Lagrangian path of the free vortex sheet, the circulation $\gamma(\alpha)$ on an infinitesimal segment is conserved, 
\begin{equation}\label{eq:freesheetGamma}
D_t(\bar{\gamma}_{k}(s,t)	s_{\alpha})=D_t\gamma_{k}(\alpha)=0,\quad k=1,2.
\end{equation}

The bound vortex sheet position, i.e. the wing position 
\begin{equation}\label{Eq:zeta_k}
\zeta_{k}(s,t)  =  X_k(t)+s+iy_k(t),\quad s\in[-1,1], 
\end{equation}
is governed by Eq.~(\ref{Eq:MX_k}). Using Eq.~(\ref{eq:w_kst}), the no-penetration boundary condition of Eq. (\ref{eq:bc}) yields
\begin{equation}\label{Eq:complex bc}
\Imag\left(w_{k}(s,t)\right)=\dot{y}_k(t),\quad s\in[-1,1],
\end{equation}
which is a Fredholm integral equation of the first kind for the bound sheet strength $\bar{\gamma}_k(s,t),\ s\in[-1,1]$.

The unsteady Kutta condition connects the bound and free sheets at the trailing edge \citep{Jones2003separatedflowinviscid}. It requires the vortex shedding to be tangential, i.e. the interface of the bound and free vortex sheets is smooth, and also determines the amount of circulation shed from the bound sheet into the free sheet:
\begin{equation}\label{eq:kuttaShedding}
\dot{\Gamma}_k(t)+\left[\Real(w_k(1,t))-U_k(t)\right]\bar{\gamma}_k(1,t)=0,
\end{equation}
where $\Gamma_k(t) = \int_{-1}^1\bar{\gamma}_k(s,t)\,\d s$ denotes the total circulation on the bound sheet, and $U_k(t)=\dot{X}_{k}(t)$ is the wing horizontal swimming speed. Kelvin's circulation conservation theorem ensures that $\Gamma_k(t) = -\int_1^{s_{\max}^k}\bar{\gamma}_k(s,t)\,\d s$.

An equation for the pressure jump $[p]_{k}(s,t)$ distributed on the bound sheet can be derived from the Euler equation (Eq.~(\ref{Eq:Euler})) \citep{Saffman1993VortexDynamics, Jones2003separatedflowinviscid, Alben2009Simulatingdynamicsflexible} and the Kutta condition Eq.~(\ref{eq:kuttaShedding}):
\begin{equation}\label{eq:pressureDist}
[p]_{k}(s,t)=\int_{1}^{s}\partial_{t}\bar{\gamma}_{k}(s',t)\,\d s'+\left[\Real(w_k(s,t))-U_k(t)\right]\bar{\gamma}_{k}(s,t)+\dot{\Gamma}_{k}(t).
\end{equation}
Note that the pressure jump across the wing contributes no force in the wing swimming direction, as the wing maintains a horizontal alignment at all times. The thrust arises from the suction force at the wing leading edge \citep{Saffman1993VortexDynamics},
\begin{equation}\label{Eq:T_kt}
T_k(t) = \frac{\pi}{8}\nu_k^{2}(-1,t),
\end{equation}
where $\nu_k(s,t)=\bar{\gamma}_k(s,t)\sqrt{1-s^2}$ has a finite value at $s=-1$ as $\bar{\gamma}_k$ has an inverse square root singularity at the wing leading edge. Note that Eq.~(\ref{Eq:T_kt}) relates the hydrodynamics with the wing dynamics through Eq.~(\ref{Eq:MX_k}).

On each wing, the instantaneous input power $P_{k,\iin}$ is defined as the product of flapping speed $\dot{y}_k$ and the pressure jump $\int_{-1}^{1}[p]_k\,\d s$, which reflects the energy required to maintain the prescribed flapping motion. The instantaneous output power $P_{k,\oout}$ equals the rate of work done by the hydrodynamic thrust $T_k$ on the wing at swimming speed $U_k$. Hence,
\begin{eqnarray}
P_{k,\iin}(t)&=&-\dot{y}_k(t)\int_{-1}^{1}[p]_k(s,t)\,\d s,\label{eq:pin}\\
P_{k,\oout}(t)&=&T_k(t)U_k(t)=\frac{\pi}{8}\nu_k^{2}(-1,t)U_k(t).\label{eq:pout}
\end{eqnarray}
The ratio of $P_{k,\iin}$ and $P_{k,\oout}$ gives the Froude efficiency on each wing $\eta_k,\ k=1,2$ and also the efficiency for the pair $\eta$ \citep{Lighthill1960Noteswimmingslender, Wu1961Swimmingwavingplate}:
\begin{equation}\label{eq:eta_k}
\eta_k=\frac{\langle P_{k,\oout}\rangle}{\langle P_{k,\iin}\rangle},\quad \eta=\frac{\langle P_{1,\oout}\rangle+\langle P_{2,\oout}\rangle}{\langle P_{1,\iin}\rangle+\langle P_{2,\iin}\rangle},
\end{equation}
where $\langle \cdot\rangle=\int_t^{t+1}(\cdot) \,\d t'$ denotes the time average over a flapping stroke.

In the end, we express the mean tangential flow speed along the wing $\overline{U}_{k,\pm}(t)$ (Eq.~(\ref{Eq:U_kpm})) using vortex sheet variables in order to calculate the skin friction drag $D_k$ of Eq.~(\ref{Eq:TkDk}). By the definitions of the mean velocity $w_k(s,t)$ and vortex sheet strength $\bar{\gamma}_k(s,t)$, the tangential speed along the two surfaces of each wing $u_{x,\pm}(\zeta_k(s,t))$ satisfies
\begin{eqnarray*}
u_{x,+}(\zeta_k(s,t))+u_{x,-}(\zeta_k(s,t)) & = & 2\Real(w_k(s,t)),\\
u_{x,+}(\zeta_k(s,t))-u_{x,-}(\zeta_k(s,t))  & = & \bar{\gamma}_k(s,t),
\end{eqnarray*}
from which we obtain
\begin{equation}\label{Eq:u_xpm}
u_{x,\pm}(\zeta_k(s,t)) = \Real(w_k(s,t))\pm\frac{1}{2}\bar{\gamma}_k(s,t).
\end{equation}
Substituting Eq.~(\ref{Eq:u_xpm}) into Eq.~(\ref{Eq:U_kpm}), $\overline{U}_{k,\pm}(t)$ is then expressed as
\begin{equation}\label{Eq:overlineU}
\overline{U}_{k,\pm}(t)=\frac{1}{2}\int_{-1}^{1}\Real(w_k(s,t))\,\d s-U_k(t)\pm\frac{1}{4}\int_{-1}^{1}\bar{\gamma}_k(s,t)\,\d s.
\end{equation}

\subsection{Numerical schemes}
Our numerical schemes follow closely those described by \citet{Fang2016computationalmodelflight} and build on similar methods that have been used to study single slender bodies undergoing unsteady motions at high Reynolds numbers \citep{Jones2005Fallingcards,Alben2008Flappingstatesflag,Alben2009Simulatingdynamicsflexible}. For the two-body interaction problem studied here, the follower's body interacts with the vortex sheet shed by the leader. To prevent the sheet from penetrating the wing due to numerical errors, an accurate quadrature rule is required. In our simulations, we improve the numerical vortex sheet method developed by \citet{Alben2009Simulatingdynamicsflexible,Alben2010Regularizingvortexsheet,Fang2016computationalmodelflight} by applying a singularity subtraction method for the singular integral kernel on the bound vortex sheet and a piece-wise accurate quadrature on the free sheets (see Appendix~\ref{sec:appendix}). Moreover, the vortex sheet scheme is combined with the skin-friction computation in our simulations to permit the study of steady-state motions that arise from force balance. In what follows, we outline the main steps of the numerical schemes.

We assume the fluid is initially at rest. The wings are initialized with horizontal velocity $U_1(0)=U_2(0)=U_0$ and wing-wing separation $x_2(0)-x_1(0)=d_0$, where $d_0=d(0)$, and $d(t)$ is defined as the horizontal separation between equivalent points on the two wings (see figure~\ref{fig:set_up}). The vortex sheet strength is initially zero, and the free sheets are confined to the wing trailing edges, i.e. $\bar{\gamma}_k(s,t)|_{t=0}=0,\ s\in[-1,1]$ and $s^k_{\max}|_{t=0}=1$. After initialization, at each time-step we update the vortex sheets in two steps:
(a) Update the existing free sheets $(\zeta_{f,k},\bar{\gamma}_{f,k})$ by an explicit scheme; 
(b) Solve and update the bound sheets $(X_k,\bar{\gamma}_{b,k})$ and wing circulation $\Gamma_k$ through an implicit nonlinear Broyden's solver \citep{Broyden1965classmethodssolving}, and shed a vortex sheet segment at the trailing edge into the free vortex sheet. Here the subscripts $f$ and $b$ denote the free and bound sheets, respectively. 

The free vortex sheets are discretized by Lagrangian points $\zeta_{f,k}^{j,n},\ 0\le j\le n$ at time $t_n$. At each time-step $t_j$, a new vortex sheet segment $[\zeta_{f,k}^{j-1,j}, \zeta_{f,k}^{j,j}]$ is produced at each wing trailing edge (no segment shed at $t_0$). The dynamics of the free sheet follows Eq.~\eqref{eq:freesheet} and is numerically updated by a 2nd-order explicit Adam-Bashforth method, except for the last segment which is newly generated at the last time-step and is updated using an explicit Euler scheme. Based on Eq.~(\ref{eq:freesheetGamma}), the vortex circulation on each vortex sheet segment is conserved. Following this, on segment $[\zeta_{f,k}^{j-1,n}, \zeta_{f,k}^{j,n}]$ the true vortex sheet strength (circulation density) $\bar{\gamma}_{f,k}^{j,n}, 1\le j\le n$,  is calculated by dividing the circulation of the segment by the segment length. For the newly generated segment at time $t_{n+1}$, $[\zeta_{f,k}^{n,n+1}, \zeta_{f,k}^{n+1,n+1}]$, the last end point $\zeta_{f,k}^{n+1,n+1}$ is the same as the wing trailing edge at time $t_{n+1}$, and the circulation of the last segment is equal to $-(\Gamma_k^{n+1}-\Gamma_k^{n})$, where both the trailing edge $\zeta_{f,k}^{n+1,n+1}$ and wing circulation $\Gamma_k^{n+1}$ at time $t_{n+1}$ are solved in the implicit step next.

The bound vortex sheet, i.e. the flapping wing, is discretized using $m+1$ Chebyshev-Gauss-Lobatto nodes 
\begin{equation}\label{Eq:Lobatto}
s_{j}=\cos(\phi_{j}),\;\phi_{j}=\frac{j\pi}{m},\quad j=0,\ldots,m,
\end{equation}
and the smooth function $\nu_k(s,t)=\bar{\gamma}_k(s,t)\sqrt{1-s^{2}}$ is approximated by the $m$th order polynomial interpolated at $s_i$. Once the free sheets are updated, in the implicit step the wing position and speed $(X_k^n,U_k^n)\approx(X_k(t_n),U_k(t_n))$, bound sheet strength ${\nu_k^{i,n}\approx\nu_k(s_i,t_n)}$ and wing circulation $\Gamma_k^{n}\approx \Gamma_k(t_n)$ are solved to match the updated free vortex sheets.
The Cauchy singular integral equation Eq.~(\ref{Eq:complex bc}) and Kutta condition Eq.~(\ref{eq:kuttaShedding}) together provide $2(m+1)$ equations for the variables $\nu_k^{i,n},\ i=1,\ldots,m$ and $\Gamma_k^{n}\approx \Gamma_k(t_n)$ \citep[see][]{Golberg2013Numericalsolutionintegral,Alben2009Simulatingdynamicsflexible,Fang2016computationalmodelflight}. 
The wing dynamical variables (see Eq.~(\ref{Eq:MX_k})) are discretized by a 2nd-order Crank-Nicolson scheme, which then provides~$4$ equations for $(X_k^n,U_k^n)$:
\begin{equation}\label{eq:Crank-Nicolson}
0 = X_k^{n+1}-X_k^{n}-\frac{\triangle t}{2}(U_k^{n+1}+U_k^{n}),\quad
0  =  U_k^{n+1}-U_k^{n}-\frac{\triangle t}{2M}(F_k^{n+1}+F_k^{n}),\quad k=1,2,
\end{equation}
where $F_k^{n}\approx F_k(t_{n})=T_k(t_{n})+D_k(t_{n})$ is the horizontal component of force on the wing, calculated using Eqs.~(\ref{Eq:TkDk}), (\ref{Eq:T_kt}), and (\ref{Eq:overlineU}). This completes a nonlinear system for $2(m+1)+4$ variables for the bound vortex sheets. Using Broyden's method, the solution is found to converge to $10^{-10}$ in about $10\sim 15$ iterations.

Some additional implementation details are worth noting. First, in the Birkhoff--Rott equation (Eq.~(\ref{eq:freesheet})), a regularized kernel \citep{Krasny1986Desingularizationperiodicvortex}
\begin{equation}\label{Eq:delta}
\mathcal{K}_{\delta}(s,s')=\frac{\overline{\zeta_k(s,t)-\zeta_k(s',t)}}{|\zeta_k(s,t)-\zeta_k(s',t)|^2+\delta^2(s)}
\end{equation}
is applied on the free vortex sheets, which suppresses instabilities and resolves the ill-posedness of the free-sheet dynamics \citep{Moore1979spontaneousappearancesingularity, Shelley1992studysingularityformation}. The regularization is singular on the bound vortex sheet. To resolve discontinuities at the wing trailing edge, the smoothing function $\delta(s)$ is defined using the velocity smoothing treatment developed by \citet{Alben2010Regularizingvortexsheet}. As each time-step introduces a new vortex segment into the free sheet from each wing, the number of points on the free sheets increases linearly in time. For computational efficiency, we approximate the far-field vortex sheets using point vortices, the error introduced being of order $O(1/r)$, where $r$ is the distance to the wing. Using $\triangle t=0.01$, the number of points needed to resolve the near-field free vortex sheets is about $10^3\sim10^4$. To speed up the integral evaluations, we use an adaptive kernel-independent fast multipole method (FMM) for the regularized kernel Eq.~(\ref{Eq:delta}) \citep{Fang2016computationalmodelflight}, and a 2D FMM for the original Laplace kernel~$\mathcal{K}(s,s')=1/(\zeta_k(s,t)-\zeta_k(s',t))$ \citep{Carrier1988fastadaptivemultipole}. In the latter, we note that the interaction is very singular when the follower gets close to the free vortex sheet shed from the leader, thus accurate quadrature rules are necessary here to prevent the free sheet from penetrating the bound sheet. In our simulations, we apply a singularity subtraction method for the singular integral kernel~$\mathcal{K}(s,s')$ on the bound sheets and a piece-wise accurate quadrature on the free sheets for interactions between the sheets. Details are provided in Appendix~\ref{sec:appendix}.

\section{Simulation results} \label{sec:simulation}

\begin{figure}
\centering
\includegraphics[width=\textwidth]{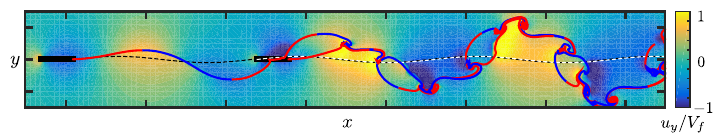}\caption{\label{fig:simulation_snapshot}{A movie snapshot from one typical simulation. Vortex sheets are shed continuously from the wing trailing edges (red for positive vortex sheet strength and blue for negative). The colormap represents the vertical component $u_y$ of flow velocity  normalized by the flapping speed $V_f=2\pi Af$. Trajectories of the leader's trailing edge (black dashed line) and follower's leading edge (white dashed line) are marked and seen to overlap.}}
\end{figure}

\subsection{Steady-state locomotion}\label{sec:steady schooling}
We begin by seeking the steady-state motions of two tandem wings with identical flapping amplitudes $\tilde{A}_1=\tilde{A}_2=\tilde{A}$. We initialize the swimming speeds $U_k$ by the stroke-averaged speed $\langle U_0\rangle$ of an isolated single wing with the same $\tilde{A}$ and drag coefficient~$C_f$, i.e. $U_1(0)=U_2(0)=\langle U_0\rangle$. Fixing $\tilde{A}=0.2$ and $C_f=0.02$, we vary the initial separation between the wings $d_0=d(0)=x_2(0)-x_1(0)$, where $d$ denotes the separation distance between the wing center points (or any equivalent points on the two wings, as in figure~\ref{fig:set_up}). As the simulation starts, the two wings interact with each other through the surrounding fluid, and the horizontal motion of each body is determined by the resulting hydrodynamic forces. The pair may eventually arrive at what seems to be a steady-state condition, and an example snapshot of the emergent configuration and flow field is shown in figure~\ref{fig:simulation_snapshot}. The pair swims or flies together at the same average speed and the two wings maintain a nearly constant separation distance.

\begin{figure}
\centering
\includegraphics[width=\textwidth]{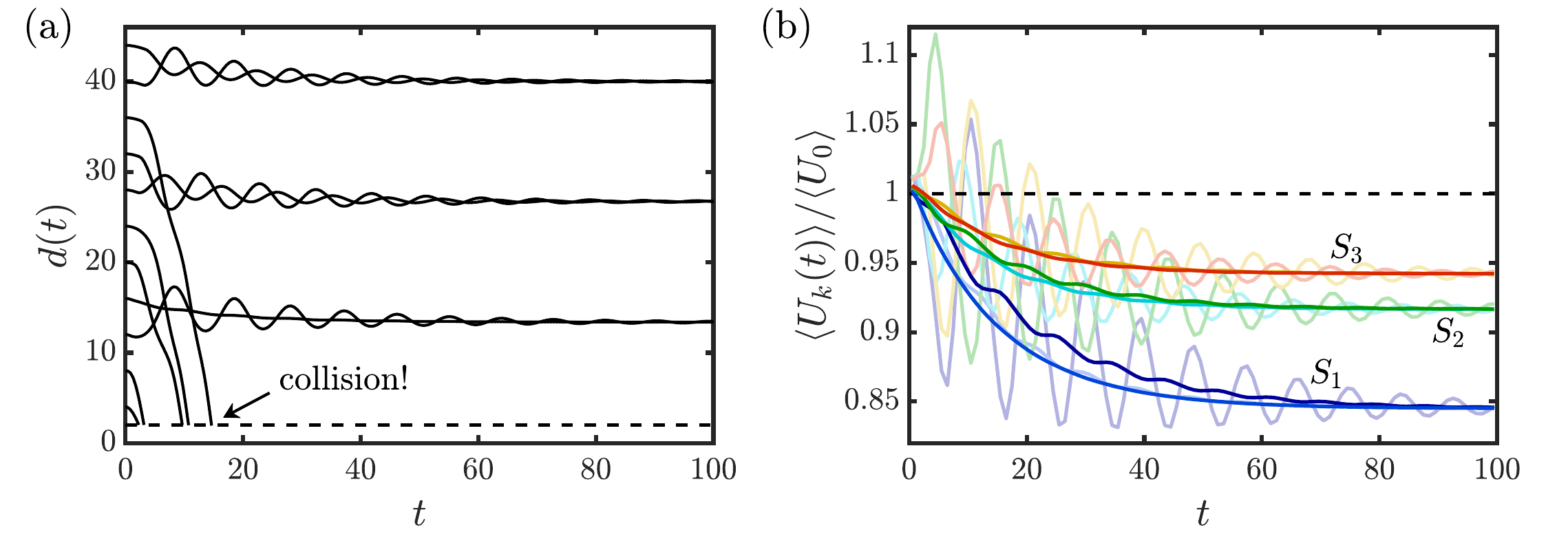}
\caption{\label{fig:fig2_approach_pdf}{Steady-state locomotion of the pair for different initial wing separations. The two wings have the same flapping amplitude $\tilde{A}=0.2$ and the drag coefficient is fixed at $C_f=0.02$. The swimming speeds of both wings are initialized by the stroke-averaged speed of an isolated single wing $\langle U_0\rangle$. Among these simulations, pairs with initial separation distances of $d_0=12,16,28,32,40,44$ are locked into `schooling states' in which the group travels together with nearly constant separation $d$. Rear-ending collisions occur for $d_0=4,8,20,24,36$. (a) Instantaneous wing separation $d(t)=x_2(t)-x_1(t)$; (b) Stroke-averaged velocity of the leader wing $\langle U_1\rangle$ (dark colors) and follower wing $\langle U_2\rangle$ (light colors), normalized by the single wing velocity $\langle U_0\rangle$. Three steady schooling states $S_1,S_2,S_3$ are displayed.
}}
\end{figure}

We examine $11$ values of $d_0 \in [4,44]$ and find several such `schooling' modes of fixed and finite steady-state separation distance $d$, as shown in figure~\ref{fig:fig2_approach_pdf}(a). Pairs initialized with $d_0=12,16, 28, 32, 40, 44$ lock onto these states, while $d_0=4,8,20,24,36$ leads to `rear-ending' collisions between the wings (and termination of the simulation). The stroke-averaged swimming velocity $\langle U_k\rangle,\ k=1,2$ is shown in figure~\ref{fig:fig2_approach_pdf}(b), which is normalized by the single wing speed $\langle U_0\rangle$. After the initialization, the leader wing slows down nearly monotonically in its approach to the terminal or steady-state velocity, while the follower's velocity $\langle U_2\rangle$ shows oscillations in its approach to this same speed.

\begin{figure}
\centering\includegraphics[width=.65\textwidth]{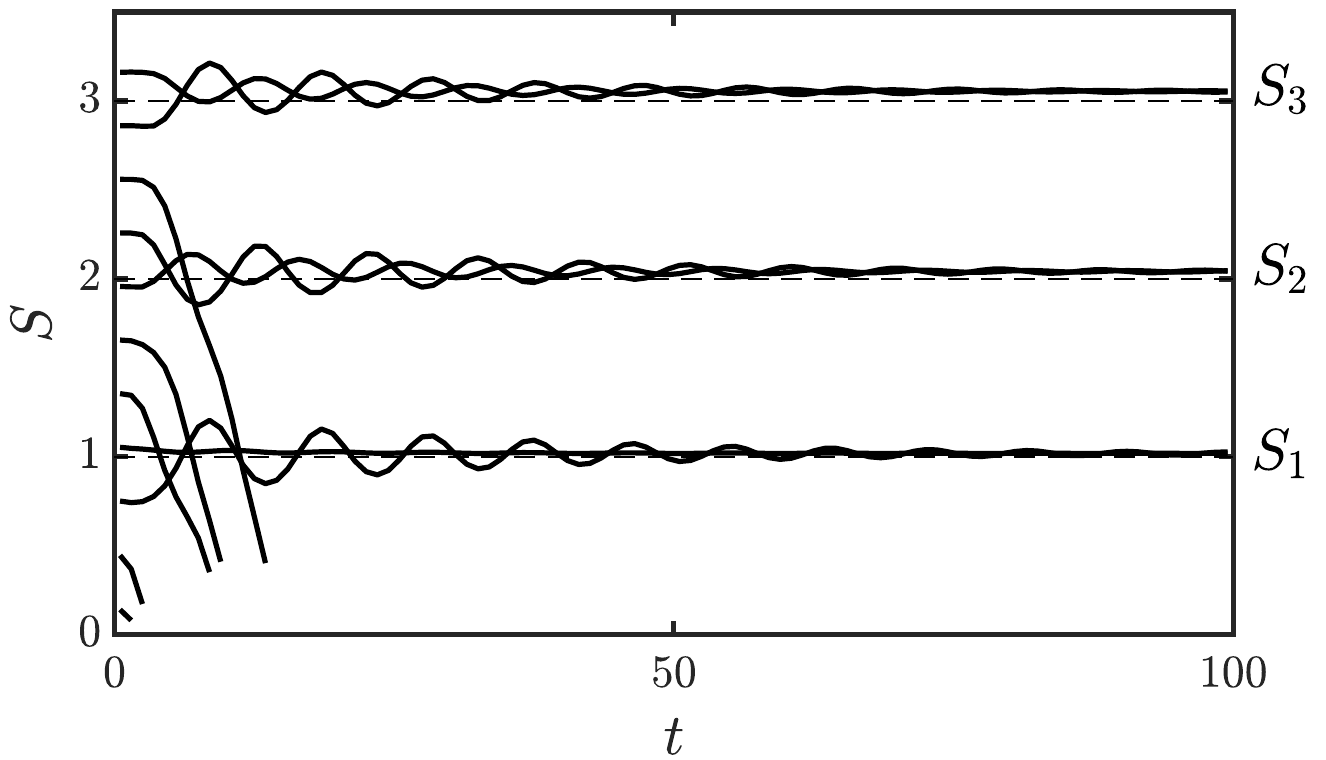}
\caption{\label{fig:fig3_sch}{Characterizing the steady-state modes using the so-called schooling number. The instantaneous schooling number is defined as $S=\langle g\rangle/\lambda $, where $\langle g\rangle =\langle d\rangle -c$ is the inter-wing gap distance and $\lambda=\langle U_1\rangle/f$ is the wavelength of the leader's path through the fluid. The steady-state or terminal values $S_1,S_2,S_3$ are close to the integers, i.e. there are an integer number of swimming wavelengths separating the two wings.}}
\end{figure}

The locomotion modes of interacting flapping wings can be quantified by the so-called schooling number \citep{Becker2015Hydrodynamicschoolingflapping,
Ramananarivo2016Flowinteractionslead,newbolt2019flow}, defined as $S=\langle g\rangle/\lambda$, which represents the stroke-averaged inter-swimmer separation $\langle g\rangle =\langle d\rangle -c$ (see figure~\ref{fig:set_up}) normalized by the wavelength $\lambda=\langle U_1\rangle/f$ of the leader's path through the fluid. Here the normalized chord length is $c=2$ and normalized flapping frequency is $f=1$. The schooling number $S$ is a dimensionless quantity that represents the relative phase of the follower's motion to the leader's. As figure~\ref{fig:fig3_sch} shows, the schooling number $S$ approaches steady-state or terminal values $S=S_1,S_2,S_3$ that are close to integer values. This indicates that the motion of the follower's leading edge is nearly in-phase spatially with that of the leader's trailing edge. This steady-state condition is made apparent in figure~\ref{fig:simulation_snapshot}, where the trajectories of the leader's trailing edge (black dashed curve) and the follower's leading edge (white) are seen to overlap. We note that more schooling states with higher values of $S$ exist and can be arrived at by considering greater initial separation distances~$d_0$. This has been observed in a recent numerical study that also employs a vortex sheet method~\citep{nitsche2024stability}. 

\begin{figure}
\centering
\includegraphics[width=\textwidth]{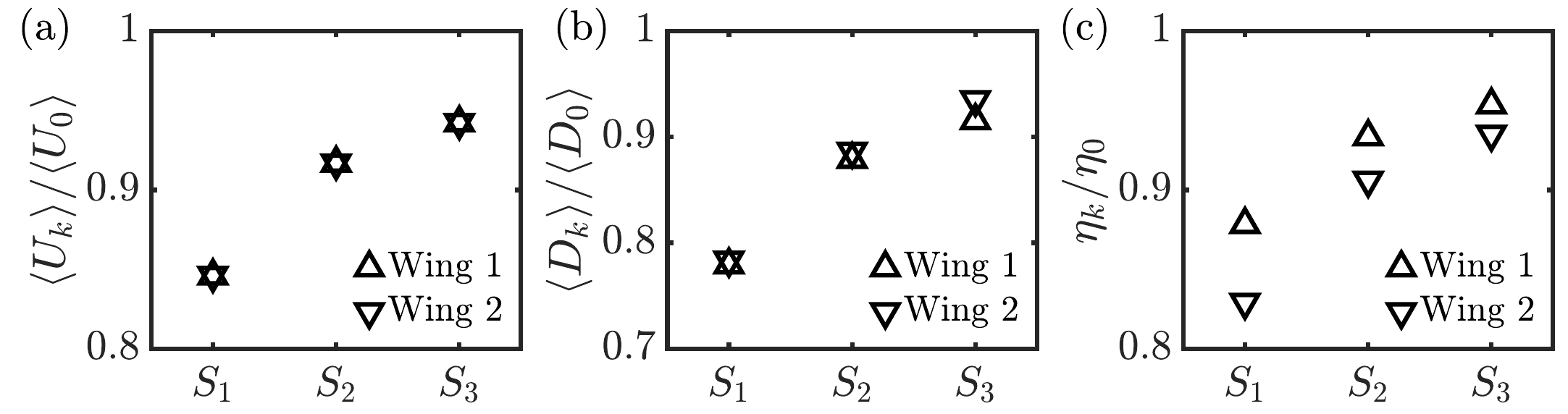}\caption{\label{fig:fig4_steady}{Comparing speed, drag and efficiency for schooling states $S=S_1,S_2,S_3$ to swimming in isolation. (a) Stroke-averaged terminal swimming speed normalized by single wing speed $\langle U_k\rangle/\langle U_0\rangle$. (b) Stroke-averaged drag at steady state normalized by single wing drag $\langle D_k\rangle/\langle D_0\rangle$. (c) Froude efficiency normalized by single wing efficiency $\eta_k/\eta_0$. }}
\end{figure}

For all the steady schooling states, we find that the terminal swimming speed is somewhat less than the single wing speed $\langle U_0\rangle$, as shown in figure~\ref{fig:fig4_steady}(a). 
The decrease in swimming speed leads to a lower skin-friction drag $\langle D_k
\rangle\apprle \langle D_0\rangle$ (figure~\ref{fig:fig4_steady}(b)), and also a lower 
efficiency $\eta_k\apprle\eta_0$ (figure~\ref{fig:fig4_steady}(c)). The speed, drag and efficiency are smaller for the lower $S$ states when the wings are closer and interactions presumably stronger, and these quantities approach the single wing values for greater $S$. 
In this respect, the interactions may be viewed as detrimental, as the two individuals would move faster and with greater efficiency if they were in isolation or were widely separated.

\subsection{Varying relevant parameters}\label{sec:vary_phy}
Next we vary some relevant model parameters and examine their effect on the first three schooling states. The flapping amplitude remains identical between the two wings and is varied over $\tilde{A}=2A/c=0.1$ to $1$, and the drag coefficient over $C_f=0.01$ to $0.1$. We note that varying $\tilde{A}$ and $C_f$ is equivalent to changing the wing kinematics, i.e. changing flapping amplitude $A$ and frequency $f$, since $C_f = 1.878\sqrt{\nu/(c^2f)}$ per Eq.~\eqref{Eq:Cf}. For each parameter pair $(\tilde{A},C_f)$, we initialize the wing speeds using the steady-state speed of the single wing with the same $(\tilde{A},C_f)$, i.e. $U_1(0)=U_2(0)=\langle U_0 \rangle$, and the separation distance is initialized such that the initial schooling numbers are integer values. 

\begin{figure}
\centering
\includegraphics[width=\textwidth]{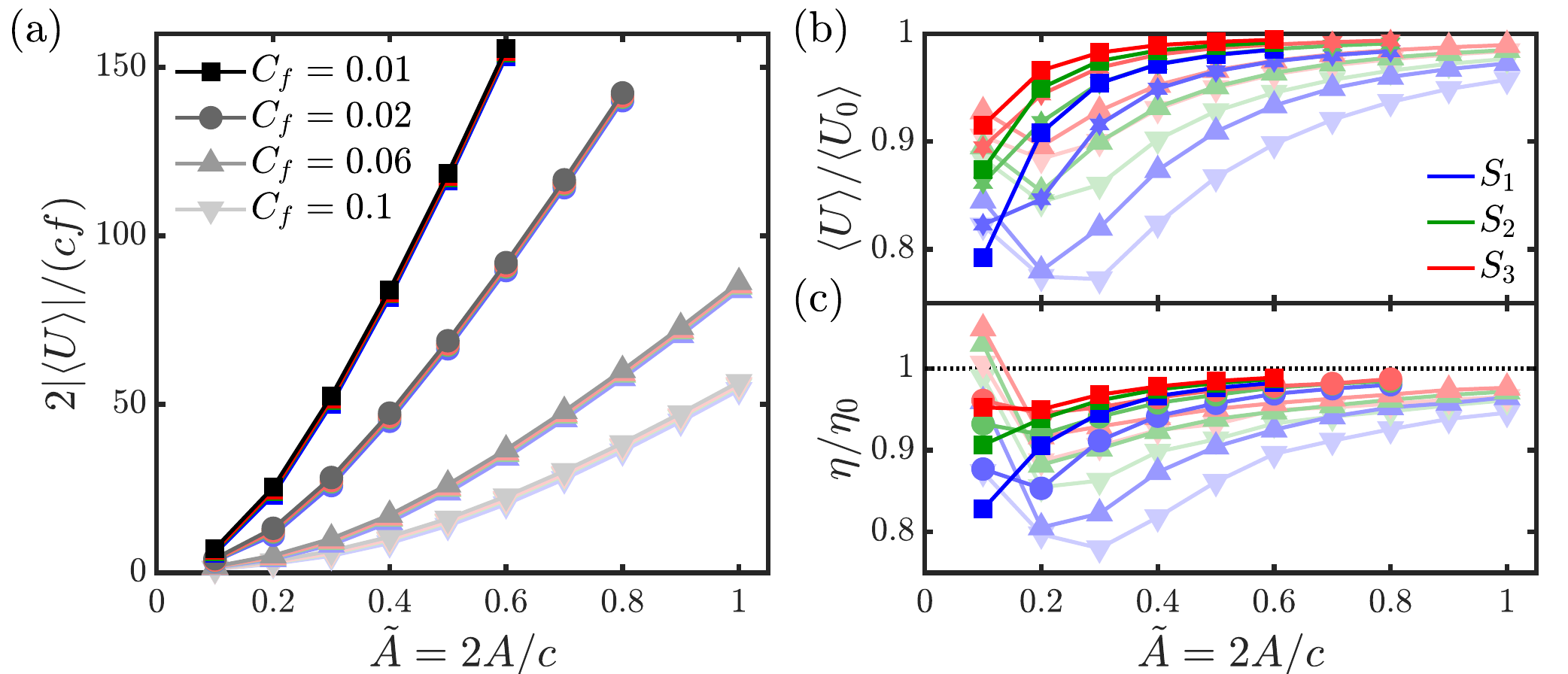}
\caption{\label{fig:school1}{Comparisons between the steady-state swimming of a pair (colored curves) and a~single wing (black and gray curves), for various flapping amplitudes $\tilde{A}$ and friction coefficients $C_f$. (a) Dimensionless steady-state and stroke-averaged swimming speed $2|\langle U \rangle|/(cf)$. (b)~Terminal swimming speed of the pair normalized by single wing speed $\langle U\rangle/\langle U_0\rangle$. (c)~Froude efficiency of the pair normalized by the single wing efficiency $\eta/ \eta_0$.}}
\end{figure}

\begin{figure}
\centering
\includegraphics[width=\textwidth]{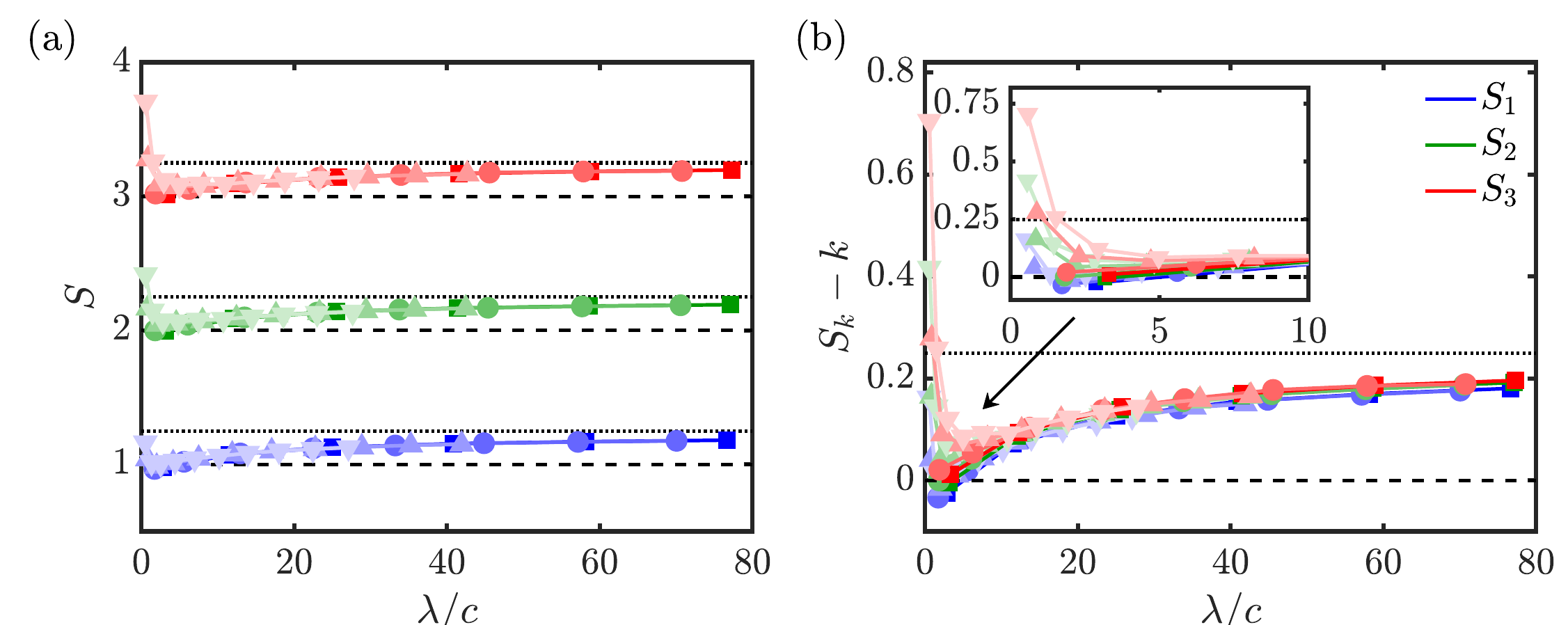}
\caption{\label{fig:school2}{Steady-state schooling numbers as functions of the normalized swimming wavelength $\lambda/c$. The flapping wing amplitude $\tilde{A}$ and drag coefficient $C_f$ are varied to arrive at different $\lambda$. (a) Three steady-state schooling numbers $S_1,S_2,S_3$ are achieved for each parameter pair $(\tilde{A},C_f)$. Except for small $\lambda/c \lesssim 2$, $S_k$ lies between integer and integer-and-a-quarter values. (b) Schooling numbers modulo the integers for states labeled $k=1,2,3$, versus the normalized swimming wavelength $\lambda/c$.}}
\end{figure}

For each parameter pair $(\tilde{A},C_f)$, we arrive at three steady states corresponding to the three initial values of $S_0 = 1,2,3$. The steady states can be characterized by the stroke-averaged terminal speed and schooling number $(\langle U \rangle, S_k)$ with $k=1,2,3$, as displayed in figure~\ref{fig:school1} and figure~\ref{fig:school2}. As shown by the black curves in figure~\ref{fig:school1}(a), the steady speed of single wing $\langle U_0 \rangle$ is a function of the wing flapping amplitude and drag coefficient $(\tilde{A},C_f)$, which is determined by the thrust and drag balance on the wing. For the same $(\tilde{A},C_f)$, the steady speed of the school $\langle U \rangle$ does not vary much from the single wing speed $\langle U_0 \rangle$. Moreover, figure~\ref{fig:school1}(b) shows that the pair speed is always slightly slower than that of an individual, with a decrease of up to $23\%$ for the parameter range we examine. The Froude efficiency of the group $\eta$ (computed using Eq.\ \eqref{eq:eta_k}), shown in figure~\ref{fig:school1}(c), is also found to be somewhat lower than the single wing efficiency $\eta_0$ for most cases. For some conditions, such as $(\tilde{A},C_f)=(0.1,0.06)$ and $(\tilde{A},C_f)=(0.1,0.1)$, the efficiency is slightly higher (up to $5\%$) than the single wing efficiency, an example of the group exploiting flow interactions to improve propulsive efficiency.

We next consider how the steady-state schooling numbers vary with changing kinematic parameters. In figure~\ref{fig:school2} we display $S$ as a function of normalized wavelength ${\lambda/c=\langle U\rangle/(cf)}$, which varies due to changes in the flapping amplitude $\tilde{A}$ and drag coefficient $C_f$. For all but the smallest wavelengths, the schooling numbers $S_k$ tend to take on approximately integer values at moderate $\lambda/c$ and monotonically increase toward integer-and-a-quarter values for large $\lambda/c$. These results show that, for sufficiently large wavelength, the motions of the follower's leading edge and the leader's trailing edge are nearly in-phase spatially, with a phase lag smaller than a quarter. This result is made more apparent in figure~\ref{fig:school2}(b), where we consider the schooling number modulo the associated integer value $S_k - k$. Comparing these results with recent experiments \citep{Ramananarivo2016Flowinteractionslead}, we see a strong correspondence in that the values of $S_k$ are close to integers for the experimentally accessed range of $\lambda/c=1$ to $8$ (see the inset of figure~\ref{fig:school2}(b)).

\begin{figure}
\centering
\includegraphics[width=\textwidth]{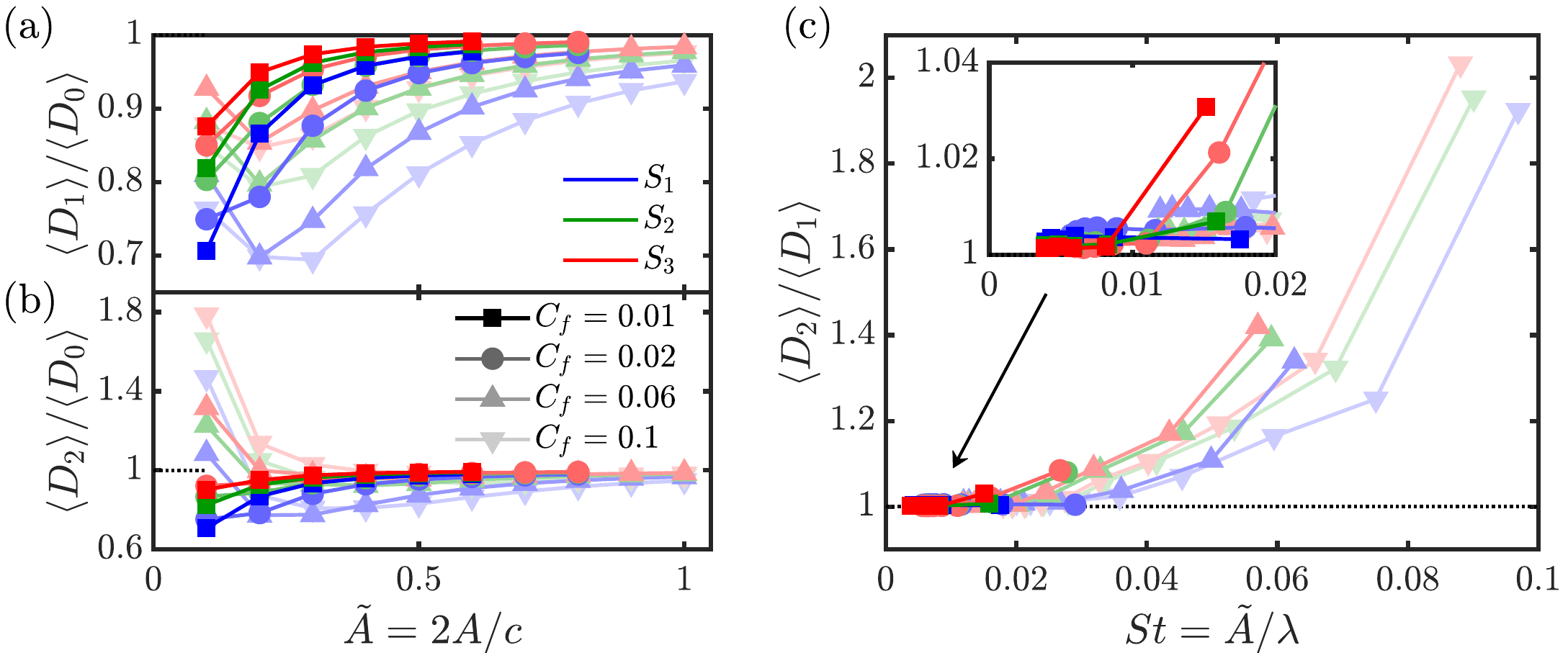}
\caption{\label{fig:school3}{(a) Drag force $\langle D_k\rangle$ experienced by (a) the leader ($k=1$) and (b) the follower ($k=2$), normalized by the single-wing drag $\langle D_0\rangle$, for various flapping amplitudes $\tilde{A}$ and friction coefficients $C_f$. (c) Follower-to-leader drag ratio, $\langle D_2 \rangle$, against the Strouhal number, $St=\tilde{A}/\lambda$.}}
\end{figure}

The drag on each of the wings is characterized in figure~\ref{fig:school3}. The leader's drag is always less than that of a single wing for the same parameters, as shown by the plot of $\langle D_1\rangle / \langle D_0\rangle$ in figure~\ref{fig:school3}(a). While it may be surprising that the leader's forces are at all affected by the presence of the follower, the drag reduction is consistent with the reduction in swimming speed $\langle U_1\rangle\apprle \langle U_0\rangle$ for the pair relative to an isolated individual (figure~\ref{fig:school1}). This result also implies, by force balance, a reduced thrust for the leader. The follower, on the other hand, may experience less or more drag compared to a single wing, depending on the values of the parameters $\tilde{A}$ and $C_f$, as shown in figure~\ref{fig:school3}(b). Cases of lower drag may again be explained by noting the reduced swimming speed of the pair relative to an individual.
Cases of higher drag arise only for high $C_f$, where we expect the higher skin friction to be balanced by higher thrust and consequently stronger wake flows. This suggests that the high drag on the follower may be related to the strong jet-like flow produced by the leader. We investigate this idea in figure~\ref{fig:school3}(c), where the follower-to-leader drag ratio is plotted against the Strouhal number, defined as $\St=Af/\langle U\rangle = \tilde{A}/ \lambda$. Low $St$ is associated with weak jet-like flows in the wake of a self-propelled flapping body -- the extreme of $St=0$ being a purely translating body that injects no momentum into the wake -- while high $St$ is associated with strong wake flows. Indeed, from figure~\ref{fig:school3}(c) we see that the follower's drag can be as high as twice that of the leader, and that this occurs only at high $St$ where we expect a strong wake generated by the leader.

\subsection{Hydrodynamic potential}\label{sec:potential}
The above results show that, as the pair approaches a steady state, the follower wing relaxes to its equilibrium position with damped oscillations (figure~\ref{fig:fig2_approach_pdf}(b)). This indicates that the hydrodynamic force on the follower acts as a restoring force that stabilizes each observed position within the leader's wake, suggesting a description in terms of a stability potential. In this view, the multiple stable positions seen in our simulations suggest a corrugated potential landscape with an array of stable wells.

\begin{figure}
\centering
\includegraphics[width=\textwidth]{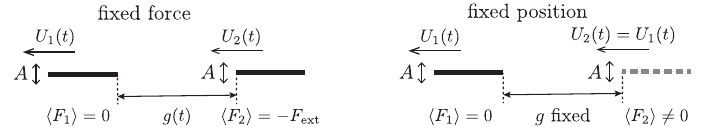}\caption{\label{fig:schematics_pot}{Two methods to map out the interaction of the follower with the leader's wake. (a) Applying an external force $F_{\mathrm{ext}}$ on the follower. At the new equilibrium, the total hydrodynamic force is zero on the leader while the hydrodynamic force on the follower is equal and opposite to the external load. (b) ``Ghost follower method.'' Fixing the position  where the follower sits relative to the leader. A fixed separation $g(t)=g_0$ is maintained by prescribing the follower dynamics to match the leader's, $U_2(t)=U_1(t)$. At the new equilibrium, the hydrodynamic thrust and drag balance on the leader but not on the follower.}}
\end{figure}

\begin{figure}
\centering
\includegraphics[width=\textwidth]{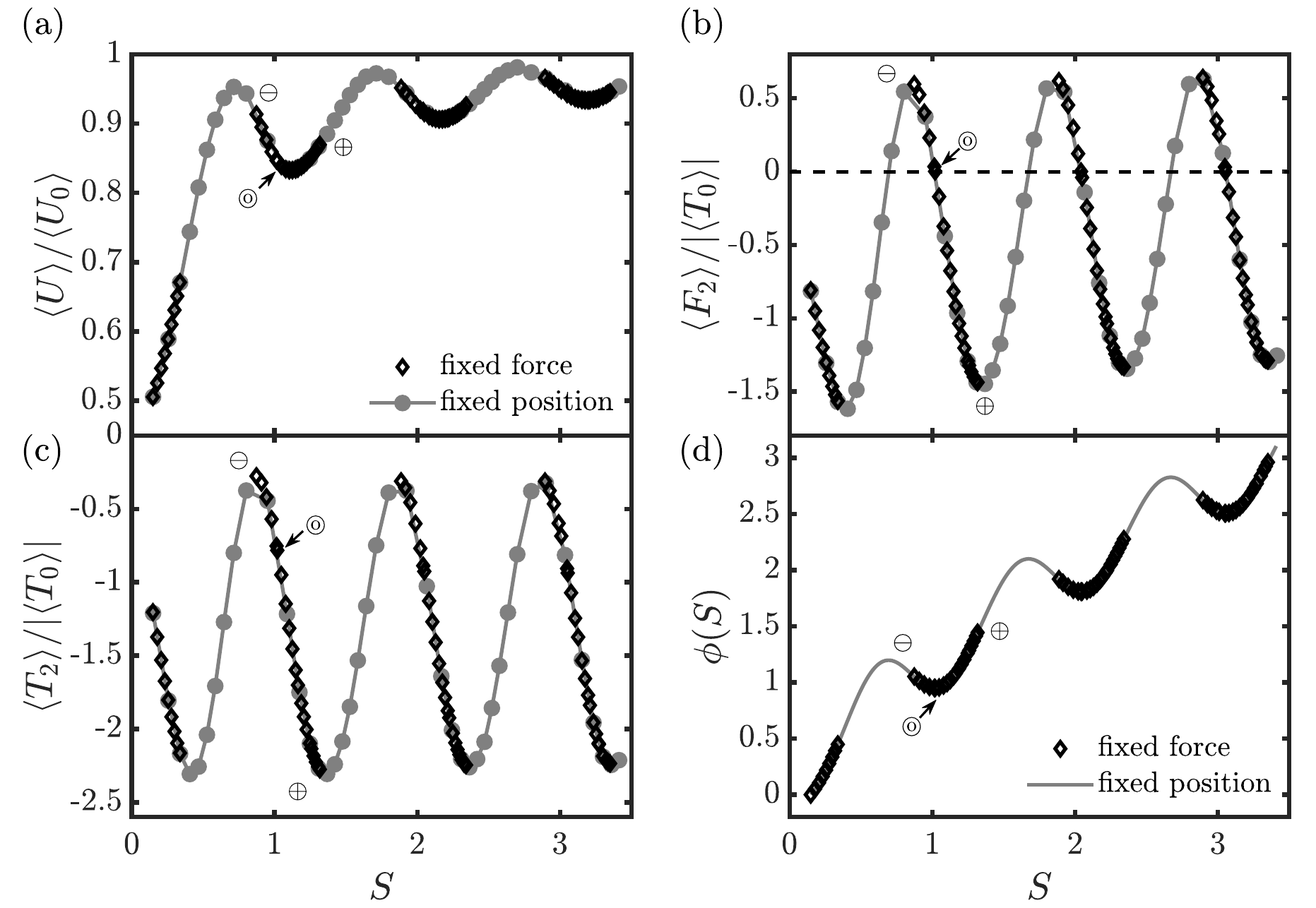}
\caption{\label{fig:potential}{Fixed-force (black open diamonds) and fixed-position (gray symbols and curves) methods for mapping out the interaction of the follower with the leader's wake. (a) Steady-state speed of the pair normalized by the single wing speed $\langle U \rangle/\langle U_0 \rangle$. (b) Total hydrodynamic force on the follower normalized by the single wing thrust $\langle F_2 \rangle/|\langle T_0 \rangle|$. (c) Thrust on the follower normalized by the single wing thrust $\langle T_2 \rangle/|\langle T_0 \rangle|$. (d) Hydrodynamic potential for the follower $\phi(S)=-\int \langle F_2\rangle \,\d S$. Three particular cases of fixed-force perturbations correspond to: zero applied force and thus an equilibrium schooling state ($\varocircle$), a negative applied force that drives the follower closer to the leader~($\varominus$), and a positive force that drives the follower away from the leader ($\varoplus$).}}
\end{figure}

To test these ideas, we map out the hydrodynamic force and potential for the follower by two methods. First, we apply a constant external load or force $F_{\mathrm{ext}}$ on the follower wing (figure~\ref{fig:schematics_pot}(a)). 
As will be shown, for sufficiently small $F_{\mathrm{ext}}$, the wings adjust their swimming dynamics and relax to new equilibrium values for the speed, gap distance, and schooling number. At the new equilibrium state, the time-averaged thrust and drag forces are balanced on each wing, and thus the total force on each is zero: for the leader $\langle F_1\rangle = 0$ and for follower $\langle F_2\rangle = -F_{\mathrm{ext}}$. At each new equilibrium obtained by varying $F_{\mathrm{ext}}$, we extract the new steady speed of the school $\langle U\rangle$, the schooling number $S$, the total force on the follower $\langle F_2\rangle$, as well as its thrust $\langle T_2\rangle$ and drag $\langle D_2\rangle$. In figure~\ref{fig:potential}(a), we report on the speed of the pair (open diamonds), which is somewhat diminished with respect to the speed of an isolated wing. Here the gaps in the data indicate that certain intervals of the schooling number $S$ are not accessed by this force perturbation procedure. In general, the pair speed tends to diminish as the two are forced closer together, but the changes in speed are not monotonic and show valleys at the free-swimming equilibria.

More importantly, we observe strong oscillations in the total force and thrust on the follower, as shown in figures~\ref{fig:potential}(b) and~\ref{fig:potential}(c), respectively. For no applied load ${F_{\mathrm{ext}} = \langle F_2\rangle {= 0}}$, we recover the equilibrium schooling states with nearly integer values of $S$, for example the point marked~$\varocircle$. Near each such state, applying a negative or leftward load on the follower tends to drive it towards the leader -- e.g. the point marked~$\varominus$ -- which reduces $S$ and which is resisted by a positive $\langle F_2\rangle$. Applying a positive load on the follower tends to drive it away from the leader -- e.g. the point marked $\varoplus$ -- which increases $S$ and which is resisted by a negative $\langle F_2\rangle$. Thus, the force on the follower due its interaction with the leader's wake is a stabilizing one. The data of figure~\ref{fig:potential}(b) also explain why certain ranges of $S$ are not observed. If $F_{\mathrm{ext}}$ is made overly large and negative, and thus $\langle F_2\rangle$ large and positive, the follower collides with the leader (i.e. $S \rightarrow 0$) rather than settling to a finite $S$. Likewise, overly large and positive $F_{\mathrm{ext}}$ leads to large and negative $\langle F_2\rangle$, and the follower separates from the leader and falls downstream (i.e. $S \rightarrow \infty$).

To map out a complete profile of the hydrodynamic restoring force, we use a second method, called the ``ghost follower method,'' in which the position of the ``ghost'' follower in the leader's wake is fixed (figure~\ref{fig:schematics_pot}(b)). Specifically, we vary the initial wing separation distance $g_0$ and prescribe this separation $g(t)=g_0$ throughout the simulation by imposing $U_2(t)=U_1(t)$. Thus, while the leader's dynamics is still determined by force balance and $\langle F_1 \rangle=0$, this is not so for the ``ghost'' follower, whose motion would necessitate a time-dependent force with a non-zero average in general, $\langle F_2 \rangle \ne 0$. In our numerical nonlinear solver (Section~\ref{sec: simulation method}), instead of solving Eq.~(\ref{eq:Crank-Nicolson}) for variables $X_2$ and $U_2$, we update the position of the follower wing at time $t_{n+1}$ by 
\begin{equation}
X_2^{n+1}=X_1^{n+1}+d_0,\quad U_2^{n+1}=U_1^{n+1},
\end{equation}
where $d_0=g_0+c$ is the initial separation distance between the wings. For a given $g_0$, the system arrives at a new equilibrium, and we measure the stroke-averaged speed $\langle U\rangle$, the total hydrodynamic force $\langle F_2\rangle$ on the follower, and its thrust $\langle T_2\rangle$, and these data are displayed in figures~\ref{fig:potential}(a)--(c) as the gray dots connected by the gray curve. The results from the two methods are seen to agree well over the intervals of $S$ accessible by both, with slight differences attributable to differences in the swimming motion within the flapping stroke. In addition, this fixed-position method accesses the gaps in the fixed-force method, which correspond to unstable conditions, as argued below.

Analogously to other mechanical systems, the potential experienced by the follower can be defined as the integral of the hydrodynamic force with respect to distance downstream from the leader. In dimensionless form, this hydrodynamic potential is $\phi(S)=-\int \langle F_2\rangle \,\d S$. In figure~\ref{fig:potential}(d) we display as the gray curve the potential as integrated from the data obtained by the fixed-position method and with cubic spline interpolation. The fixed-force method (open diamond symbols) yields a similar profile over the accessible intervals of $S$. A spline interpolation for the total hydrodynamic force is used to fill in the forbidden gaps and obtain the corresponding potential. Both methods reveal a sloped and corrugated potential (tilted washboard) whose form helps to explain the key results from the simulations. The local minima near integer $S$ are stable wells and correspond to the schooling modes, which may be arrived at by initializing the pair sufficiently close to the associated steady-state separation distance. If initialized at distances corresponding to the peaks, the follower may `fall' down the potential landscape to the global minimum of $S \approx 0$, explaining the collisions that occur for some values of $d_0$ (see figure~\ref{fig:fig2_approach_pdf}(a)). Further, the gaps in the fixed-force method can be seen to correspond to unstable conditions in which the potential has downward concavity, i.e. $\d^2 \phi/\d S^2 < 0$.

\begin{figure}
\centering
\includegraphics[width=\textwidth]{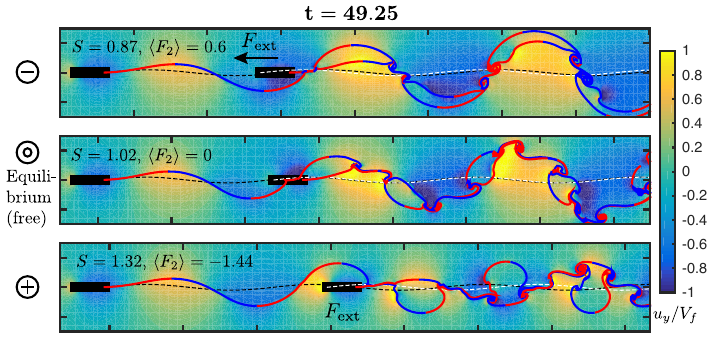}\caption{\label{fig:potential_wake}{Flow fields for a wing pair with different forces applied to the follower. The snapshots are taken during the middle of the downstroke. Vortex sheets are shed continuously from the trailing edges, with red for positive vortex sheet strength and blue for negative. The background color represents the vertical component of flow velocity $u_y$ normalized by wing flapping speed $V_f=2\pi Af$. Trajectories of leader's trailing edge (black dashed line) and follower's leading edge (white dashed line) are also shown. For the case of minimal thrust generated by the follower ($\varominus$), the follower flaps downward in a downward flow of the leader's wake. For the case of maximal thrust generated by the follower ($\varoplus$), the follower flaps downward in an upward flow. For the equilibrium schooling state ($\varocircle$, $S \approx 1$), the follower spans what would be a node in the leader's wake, with downward flows ahead and upward flows behind.}}
\end{figure}

\subsection{Flow fields and stabilization mechanism}\label{sec:flowvis}
To better understand the origin of the stable schooling states, we next compare the flow fields generated by the pair for different applied forces on the follower and thus different interactions with the leader's wake. In figure~\ref{fig:potential_wake} we show the collective wake structures from simulations corresponding to the three conditions marked in figure~\ref{fig:potential}: decreased separation distance and decreased thrust generated by the follower ($\varominus$); a schooling state equilibrium ($\varocircle$); and increased separation and increased thrust for the follower ($\varoplus$). 
In this context, increased thrust refers to a more negative thrust, as the propulsive thrust is considered negative due to the wings swimming from right to left.
The decreased and increased thrust cases correspond to the ends of the stable branch spanning $S=1$ in  figure~\ref{fig:potential}. In these snapshots of the steady-state or terminal condition, the wings are captured at mid-stroke and flapping downward, and the color map represents the vertical component of the flow $u_y$ normalized by flapping speed of the wing $V_f=2\pi Af$. When the follower's thrust reaches its minimum in the direction of swimming ($\varominus$), we find that the follower wing flaps downward in the downward flow of the leader's wake, i.e. the follower's flapping motion is in-phase with the wavy wake flow it encounters. This constructive interaction reinforces the flow, leading to a collective wake with large regions of lateral flow. For the maximum thrust case ($\varoplus$), however, the follower wing flaps downward in the upward flow of the leader's wake, i.e. the follower's flapping motion is out-of-phase with the leader's wavy wake. This destructive interaction leads to a narrower collective wake with weaker flows. The equilibrium schooling state ($\varocircle$) is intermediate between these two extremes, with a collective wake that is more similar to that of a single swimmer in its structure and strength.

This flow field characterization suggests a simple mechanism for the stability of the equilibrium schooling states. A perturbation that decreases the separation distance of the follower from the leader ($\varominus$) leads to reduced thrust since the follower flaps in-phase with the flow and thus experiences decreased vertical velocity \emph{relative to the fluid}. Reduced thrust tends to slow the follower and thus restore it towards the equilibrium schooling state position. A perturbation that increases the separation distance ($\varoplus$) leads to increased thrust since the follower flaps out-of-phase with the flow and thus experiences increased vertical velocity relative to the fluid. Enhanced thrust tends to accelerate the follower towards the leader and again restores the equilibrium schooling state position. Thus it seems the foil-wake interaction can be understood by considering the relative flow during motion through a spatially-varying flow field. These results also show that thrust may be enhanced by appropriate positioning within a wake flow, an effect that we exploit in what follows.

\subsection{A `freeloading' follower with reduced flapping amplitude}

The perturbation studies discussed in Section~\ref{sec:potential} show that the follower wing may experience higher or lower thrust depending on its relative position in the wake flow generated by the leader. To explore how wake interactions might be exploited, we consider the case of a `lazy' and an `overpowered' follower that flaps at reduced and increased amplitude relative the leader, respectively. In the absence of interactions, reduced amplitude would of course lead to slower swimming \citep{Vandenberghe2004Symmetrybreakingleads}, but low-amplitude swimming in a wake flow presents the opportunity to situate at favorable locations where the follower-flow relative velocity may still be high. To study this `freeloading' follower problem, we fix the flapping amplitude of the leader wing at $A_1=0.2$, vary the follower's amplitude $A_2$, and seek the first steady schooling state near $S=1$. Fixing $C_f=0.02$, we find steady states for ${\varepsilon=A_2/A_1\in[0.4,1.3]}$ in which the two wings `school' together despite their dissimilar kinematics. For low amplitudes $\varepsilon\apprle 0.4$, the follower drifts downstream and is unable to generate sufficient thrust to keep pace with the leader. For $\varepsilon\apprge 1.3$, the follower generates high thrust that leads to a rear-ending collision with the leader.

\begin{figure}
\centering
\includegraphics[width=\textwidth]{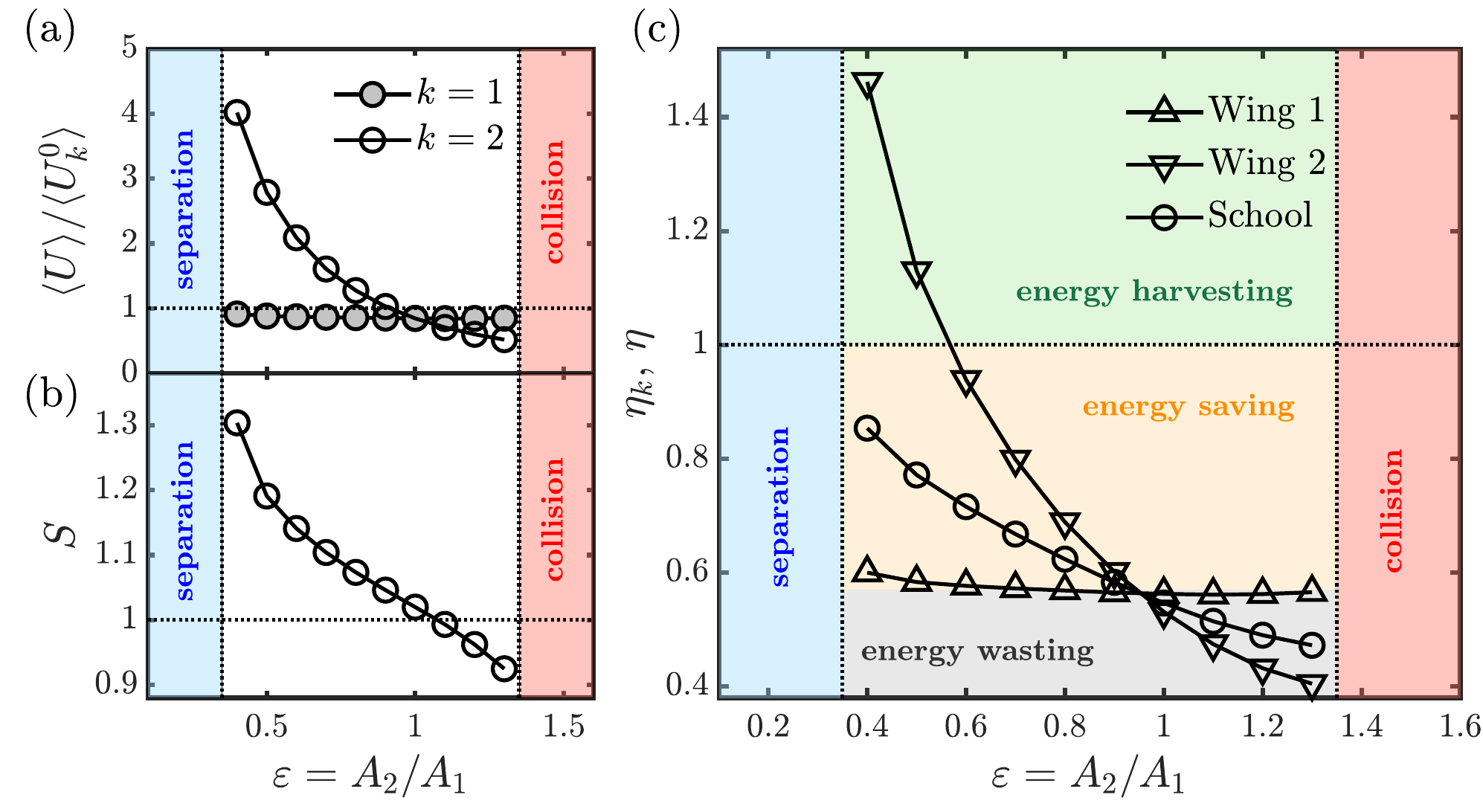}
\caption{\label{fig:bigsmall}{Speed, schooling number, and efficiency for a wing pair in which the follower has reduced or increased flapping amplitude relative to the leader. The leader's amplitude is fixed at $A_1=0.2$ and the follower's $A_2$ is varied. Steady schooling states near the $S=1$ branch are found for $\varepsilon=A_2/A_1\in[0.4,1.3]$. For $\varepsilon\apprle 0.4$, the follower can separate from the leader, while for $\varepsilon\apprge 1.3$, the follower can collide with the leader. (a) Swimming speed $\langle U\rangle$ of the pair normalized by the speed of a single wing having the leader's amplitude $\langle U_{1}^{0}\rangle$ (gray circles) or follower's amplitude $\langle U_{2}^{0}\rangle$ (empty circles). (b) First steady schooling state $S_1$. (c) Froude efficiency of the two wings $\eta_1$ and $\eta_2$, and efficiency of the school $\eta$, as given by Eq.~\eqref{eq:eta_k}. The leader's efficiency is nearly unchanged but the follower experiences a strong increase in efficiency for $\varepsilon<1$. The follower's efficiency may even exceed unity, indicating that a `lazy' follower can effectively extract energy from the leader's wake. The school exhibits improved efficiency when $\varepsilon<1$ but remains below unity, saving energy through flow interactions. An `overpowered' follower ($\varepsilon>1$) causes energy to be wasted, as its efficiency falls below the level it would achieve in isolation.}}
\end{figure}

The steady schooling states are characterized by the speed, schooling number, and efficiency plots of figure~\ref{fig:bigsmall}. As the follower's amplitude $A_2$ is varied, we find the school speed $\langle U\rangle$ remains somewhat smaller than the speed $\langle U_{1}^{0}\rangle$ of a single wing with the leader's flapping amplitude $A_1$, as shown by the gray circles in figure~\ref{fig:bigsmall}(a). 
However, if the pair speed is compared to a single swimmer of the follower's amplitude $\langle U_{2}^{0}\rangle$, we observe a speed enhancement for $\varepsilon < 1$, as shown by the empty circles in figure~\ref{fig:bigsmall}(a). This implies that the follower swims at higher speed than it would in isolation, with a speed enhancement of up to four times. This exploitation of the wake flow for $\varepsilon < 1$ is associated with higher schooling numbers $S$, as shown in figure~\ref{fig:bigsmall}(b). The flow field analysis given above suggests an interpretation of these findings: The weakly-flapping follower re-locates to a downstream position in which the wake flow is out-of-phase with the flapping motion, thereby maintaining strong relative flows and sufficiently high thrust to keep pace with the leader. Increasing $\varepsilon$, which corresponds to an `overpowered' follower, leads to smaller $S$. This is explained by the increase in the thrust associated with faster flapping of an isolated single wing, which is counteracted by a compensating decrease in thrust associated with moving closer to the leader, where the follower’s upstroke occurs in a stronger upward flow~\citep{newbolt2019flow}. 
 
Finally, we also consider the efficiency of each wing and the pair, as shown in figure~\ref{fig:bigsmall}(c). Because the swimming speed of the pair is nearly unchanged by varying $\varepsilon$, the output power of the leader and follower is nearly unchanged. This leads to little change in the Froude efficiency of the leader. The follower, on the other hand, experiences a strong increase in efficiency for $\varepsilon < 1$ since its input power decreases with decreasing flapping amplitude. The efficiency of the follower may even exceed unity ($\eta_2>1$) when $\varepsilon<0.6$, another manifestation of how wake effects may be exploited. Essentially, a `lazy' follower can extract more energy from the wake of the leader than it is putting in, getting a ``boost'' from the leader's wake. 
The school also exhibits improved efficiency when $\varepsilon<1$, although
$\eta$ remains below unity. This indicates that, while the school still performs work when the follower is `lazy', its overall performance is improved compared to when the wings are operating independently, saving energy through the flow interactions. In contrast, an `overpowered' follower ($\varepsilon>1$) may experience reduced efficiency compared to what it could achieve in isolation. In this case, the flow interactions actually lead to wasted energy.

\section{Linearized model}\label{sec:linearTheory}
To understand the steady schooling of the two wings and the hydrodynamic potential on the follower wing, we adapt a small-amplitude linearized theory \citep{Wu1961Swimmingwavingplate,Wu1975Extractionflowenergy} to our schooling problem of two wings in tandem.
The wing is modeled by a flat rigid plate of chord length $c$, moving with horizontal speed $U_{\infty}$ (in the negative $x$-direction) with prescribed vertical motion $h(t)=A\cos(2\pi f t)$, in a two-dimensional incompressible inviscid fluid. Assuming small flapping amplitude $A/c \ll 1$ and small Strouhal number $\St=Af /U_{\infty}\ll 1$, the 2D Euler equation as well as the ``no-penetration" boundary condition are linearized and solved at the wing boundary using Fourier series and conformal mapping. The thrust and efficiency can thus be calculated from the solution.

We first adapt the linearized theory developed by \citet{Wu1961Swimmingwavingplate} for the swimming of a waving plate, to our case of a heaving plate swimming in a quiescent  background. It models the swimming of an isolated single wing, and also serves as a good approximation of the leader wing in the school. This approximation is motivated by experiments \citep{Ramananarivo2016Flowinteractionslead} and our simulations, both of which show the leader wing is affected weakly by the follower, and the school speed differs from the single wing speed by less than $20\%$ regardless of the amplitude and location of the follower wing (see figures~\ref{fig:school1}(b), \ref{fig:potential}(a) and \ref{fig:bigsmall}(a)). Combining the thrust formula provided by Wu's solution with the Blasius skin-friction formula Eq.~(\ref{Eq:Df}), the free swimming speed of the single wing, or the leader wing, can be determined. 

Under assumptions of small amplitude $A/c\ll 1$ and small Strouhal number $\St\ll 1$, the vortex sheet generated by the leader wing can be linearized as a flat vortex sheet, the flow induced by which is then approximated by a simple steady stationary wave. We then examine the interaction of the follower wing with this wave by adapting a similar linearized theory of \citet{Wu1975Extractionflowenergy}, in which an oscillating plate in a wavy background stream is solved. Provided by the solution, the thrust of the follower wing can be expressed as a function of the schooling number $S$.
The drag is also provided by the Blasius skin-friction formula Eq.~(\ref{Eq:Df}), using the swimming speed as the boundary-layer stream speed, assuming the follower does not feel an extra fluid jet produced by the leader. Combining the thrust and the drag, we can then calculate the hydrodynamic potential on the follower and solve the steady schooling numbers analytically.

\subsection{Linearized model for isolated flapping wings}


\subsubsection{Swimming of a single flapping wing} \label{sec: singleQues}

We begin by considering a single wing swimming with constant speed $U_{\infty}$ in a quiescent background.
We use the same non-dimensionalization as in Section~\ref{sec: simulation method}: the characteristic length $\tilde{L}=c/2,$ time $\tilde{T}=1/f,$ velocity $\tilde{U}=cf/2$
and pressure $P=\rho_f\tilde{U}^2=\rho_{f}c^{2}f^{2}/4$. The heaving motion of the wing described in the body frame is,
\begin{equation}
h(x,t)=A\cos(2\pi t),\quad -1\le x\le 1,
\end{equation}
where the wing spans from $x=-1$ to $x=1$. 

In the frame of the flow, we assume the wing translates with constant velocity $-U_{\infty}$ in the negative $x$-direction. Then in the wing frame, the wing is considered to be flapping in a uniform background stream of constant velocity $U_{\infty}$ in the positive $x$-direction. Following \citet{Wu1961Swimmingwavingplate}, the flow velocity is denoted as $\mathbf{q}=(U_{\infty}+u,v),$  where velocity perturbations $(u,v)$ are assumed small compared to the uniform stream $U_{\infty}$, i.e. $U_{\infty}\gg u$ and $U_{\infty}\gg v$. The flow satisfies the 2D Euler equation, and is linearized as,
\begin{equation}\label{Eq:linearEuler}
\partial_{t}\mathbf{q}+U_{\infty}\partial_x \mathbf{q} = -\nabla p,\quad
\nabla\cdot\mathbf{q} = 0.
\end{equation}
The kinematic boundary condition on the wing requires the velocity to be continuous in the normal direction. Assuming the heaving amplitude is small, $A \ll 1$, the  boundary condition is then linearized as
\begin{equation}\label{Eq:linear boundary}
v(x,y,t)\mid_{y=\pm0}=\partial_t h(x,t) ,\quad -1\le x\le1.
\end{equation}
Note that in dimensional quantities, Eq.~(\ref{Eq:linear boundary}) and condition $U_{\infty}\gg v$ imply that the Strouhal number $\St=Af/U_{\infty}$ is small, as $U_{\infty} \gg v\sim Af$, where the Strouhal number denotes the ratio of the vertical flapping speed and the translational speed of the wing.

Instead of solving the fluid velocity directly, \citet{Wu1961Swimmingwavingplate} solves for acceleration potential
\begin{equation} \label{Eq:acceleration potential}
f(z,t) = \phi(x,y,t) + i\psi(x,y,t),
\end{equation}
in complex variables $z=x+iy$. Here,
\begin{equation}
\phi(x,y,t) = -p
\end{equation}
is Prandtl's acceleration potential, which is a harmonic function of $(x,y)$ for every $t$, and $\psi(x,y,t)$ is its harmonic conjugate. 
The pressure $p$ is continuous everywhere except across the wing boundary, $-1 \le x \le 1$. It then follows that $\phi$ is regular inside the flow, so that
\begin{equation}\label{Eq:PhiJump}
\phi(x,+0,t)=\phi(x,-0,t), \quad \text{for} \; |x|>1,
\end{equation}
and for all $t$.
The Kutta condition is required at the trailing edge of the wing,
\begin{equation}
|f(1,t)| < \infty, \quad \forall t \ge 0.
\end{equation}
Furthermore, far-field boundary conditions 
\begin{equation}\label{Eq:fardecay}
f(z,t)\rightarrow 0 \quad \text{as}\quad |z|\rightarrow \infty; \quad u(z,t)-iv(z,t)\rightarrow 0 \quad \text{as}\quad z\rightarrow -\infty,
\end{equation}
are required to complete the problem.

\citet{Wu1961Swimmingwavingplate} constructed the solution of $f(z,t)$ using Fourier series and conformal transformation. Provided by the solution, the stroke-averaged hydrodynamic thrust on the wing is given as
\begin{equation}\label{Eq:T_single}
\langle T \rangle = -4\pi^{3}A^{2}\left|C(\sigma)\right|^2,
\end{equation}
\begin{equation}\label{Eq:C}
C(\sigma)=\frac{K_{1}(i\sigma)}{K_{0}(i\sigma)+K_{1}(i\sigma)}=C_{1}(\sigma)+iC_{2}(\sigma),\quad \sigma=2\pi/U_{\infty},
\end{equation}
where $K_0$ and $K_1$ are the modified Bessel functions of the second kind of orders zero and one. $C(\sigma)$ is called Theodorsen's function, and $C_1(\sigma)$ and $C_2(\sigma)$ are the real and imaginary parts of $C(\sigma)$, respectively. Using dimensional quantities,  $\sigma = \pi cf/U_{\infty}$ is usually called the reduced frequency. If we consider the problem in the flow frame, that the wing moves with velocity $U_{\infty}$ in a quiescent background, $\lambda=U_{\infty}/f$ denotes the wavelength of the wing motion, and then $\sigma = \pi c/\lambda$ represents the ratio of the wing size to the wavelength. The thrust (Eq.~(\ref{Eq:T_single})) is in the same direction as swimming (negative $x$-direction) and is composed of the leading-edge suction only. As discussed in Section~\ref{subsec:simulation model} regarding the vortex sheet model, the pressure distributed along the wing has zero horizontal component, since the body is always horizontally aligned with the flow.

The Froude efficiency \citep{Lighthill1960Noteswimmingslender}, or the stroke-averaged efficiency of propulsion, is defined in the same way as in the vortex sheet model (Section~\ref{subsec:simulation model}), which is provided by \citet{Wu1961Swimmingwavingplate} as,
\begin{equation}\label{Eq:WuEta}
\eta = \frac{\langle P_{\mathrm{out}} \rangle}{ \langle P_{\mathrm{in}} \rangle} = \frac{\left|C(\sigma)\right|^2 }{C_{1}(\sigma)}.
\end{equation}
Here the output power
\begin{equation}
\langle P_{\mathrm{out}}(t) \rangle=\langle T(t) \rangle U_{\infty} 
\end{equation}
is computed from Eq.~(\ref{Eq:T_single}). The input power required to maintain the flapping motion is expressed as
\begin{equation}\label{Eq:WuPin}
\langle P_{\mathrm{in}} \rangle
= \biggl \langle -\int_{-1}^{1}[p] \partial_t h \,\d x \biggr \rangle;
\end{equation}
calculated by integrating the instantaneous pressure jump distribution $[p](x,t)$ along the wing, where
\begin{equation}
[p](x,t)=p(x,+0,t)-p(x,-0,t),\quad -1\le x\le 1,
\end{equation}
computed from Wu's solution of $f(z,t)$.

\begin{figure}
\centering
\includegraphics[width=\textwidth]{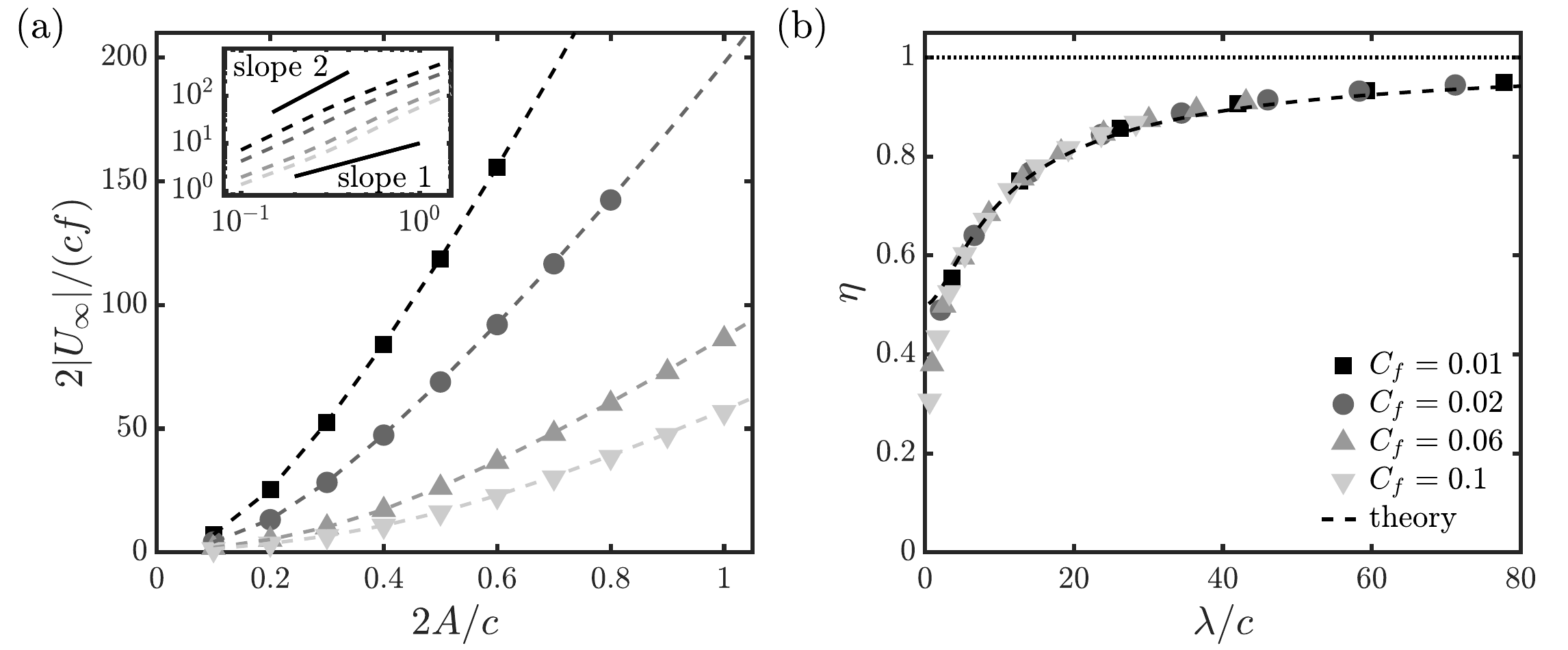}
\caption{\label{fig:singleSteady}{Comparison of linear theory (dashed lines) with simulation (symbols) of single self-propelled flapping wing, for various flapping amplitudes $A$ and drag coefficients $C_f$. Simulation data is the same as in figure~\ref{fig:school1}. (a) Steady swimming speed $|U_{\infty}|$, normalized by the characteristic speed $cf/2$. The power law of steady speed $2|U_{\infty}|/(cf)$ in the amplitude $2A/c$ is between linear and quadratic, as shown in the log--log plot in the inset. (b) The Froude efficiency $\eta$ of the single wing at steady state. The linearized theory Eq.~(\ref{Eq:WuEta}) predicts limits of $\eta\rightarrow 1$ as $\lambda/c\rightarrow +\infty$, and $\eta\rightarrow 1/2$ as $\lambda/c\rightarrow 0$.}}
\end{figure}

Note that using dimensional quantities, the thrust formula Eq.~(\ref{Eq:T_single}) is expressed as
\begin{equation}\label{Eq:dimen T_single}
\langle T \rangle = -2\pi^{3}c \rho_{f} A^{2}f^{2}\left|C(\sigma)\right|^2, \quad \sigma=\pi cf/U_{\infty},
\end{equation} 
which is quadratic in flapping velocity $V_f=2\pi Af$ for fixed reduced frequency $\sigma$. The limiting case of $\sigma\rightarrow 0$ is of particular interest. For small $\sigma$,  $K_0(i \sigma)$ and $K_1(i\sigma)$ can be expanded as 
\begin{equation}\label{Eq:K0aym}
K_{0}(i\sigma) = -\ln\sigma+\ln2-\gamma-i\pi/2 + O(\sigma^{2}\ln \sigma),
\end{equation}
\begin{equation}\label{Eq:K1aym}
K_{1}(i\sigma)  =-\frac{\pi}{4}\sigma-\frac{i}{\sigma}+\frac{i\sigma}{4}\left(2\ln \sigma-2\ln 2+2\gamma-1\right)+ O(\sigma^{3}\ln \sigma),
\end{equation}
where $\gamma\approx 0.5771$ is Euler's constant \citep{Abramowitz1964Handbookmathematicalfunctions}. Substituting Eqs.~\eqref{Eq:K0aym}--\eqref{Eq:K1aym} in Theodorsen's function Eq.~(\ref{Eq:C}), we obtain $C(\sigma)\rightarrow 1$, $C_1(\sigma)\rightarrow 1$ as $\sigma\rightarrow 0$. Hence $\langle T \rangle \sim -2\pi^{3}c \rho_{f} A^{2}f^{2}$ for small $\sigma$, or long free swimming wavelength $\lambda/c=\pi /\sigma$. 
The Froude efficiency $\eta$ in Eq.~(\ref{Eq:WuEta}) is determined only by the reduced frequency $\sigma$, and has limit $\eta \rightarrow 1$  as $\sigma\rightarrow 0$ (see figure~\ref{fig:singleSteady}(b)).
In the other limiting case of $\sigma \rightarrow +\infty,$ we could obtain $C(\sigma)=C_1(\sigma)\rightarrow 1/2$,  and thus $\langle T \rangle \sim -\pi^{3}c \rho_{f} A^{2}f^{2}$ and $\eta \rightarrow 1/2$ for $\sigma\gg 1$. 
The limit at large $\sigma$, or small free swimming wavelength $\lambda/c=\pi /\sigma$, is not of much interest to free swimming problems, as the free swimming speed $U_{\infty} = \pi cf/\sigma$, determined by the flapping amplitude $A$ and frequency $f$, does not approach a small value for the parameter ranges of our interest.

For the freely swimming wing, the vertical flapping kinematics determines the free translational speed in the horizontal direction, that is, the free speed $U_{\infty}$ is a function of the flapping amplitude $A$ and frequency $f$. The steady free speed $U_{\infty}$ can be solved from balancing the thrust and drag. The thrust $\langle T \rangle$ is provided by Eq.~(\ref{Eq:T_single}) (dimensionless), or Eq.~(\ref{Eq:dimen T_single}) (dimensional). The drag $D$ is modeled using a simplification of the skin-friction model Eq.~(\ref{Eq:TkDk}) used in simulations, where the boundary layer velocity is approximated using the free swimming velocity of the wing,
\begin{equation}
D=2C_f|U_{\infty}|^{3/2},
\end{equation}
and the drag coefficient $C_f$ is provided by Eq.~(\ref{Eq:Cf}). The dimensionless free swimming speed $U_{\infty}$ is then solved from balance of forces
\begin{equation}\label{Eq:linear balance}
0=\langle T \rangle + D = -4\pi^{3}A^{2}\left|C(\sigma)\right|^2+2C_f|U_{\infty}|^{3/2},\quad \sigma=2\pi/U_{\infty}.
\end{equation}
The steady swimming speed $U_{\infty}$, solved from Eq.~(\ref{Eq:linear balance}), is shown in figure~\ref{fig:singleSteady}(a), compared with the steady speed of single flapping wing calculated through the vortex sheet simulations (see figure~\ref{fig:school1}(a)). Figure~\ref{fig:singleSteady}(b) compares the Froude efficiency of the single wing at steady swimming state, calculated using vortex sheet simulation and the linearized theory Eq.~(\ref{Eq:WuEta}). The results from linearized theory agree well with the simulation results. For the parameter values used in our simulations, the flapping amplitude ($A/c = 0.05$ to $0.5$) and the Strouhal number ($\St\approx 0.005$ to $0.1$) are within small ranges, so the linearization provides a good approximation.
The log--log plot of the free speed, in the inset of figure~\ref{fig:singleSteady}(a), shows that the free speed $U_{\infty}$ is superlinear in flapping amplitude, which agrees with the power law in experiments. This will be discussed further in Section~\ref{sec:discussion}.

\subsubsection{Fluid wake generated by single flapping wing}

The vortex wake generated by a flapping wing is complex and self-developing, as shown by our simulation figure~\ref{fig:potential_wake} and by experimental flow visualizations 
\citep{Vandenberghe2006unidirectionalflightfree,
Ramananarivo2016Flowinteractionslead}. In the linear regime, the Strouhal number $St=Af /U_{\infty}$, the ratio of the vortex wake width $A$ and wavelength $\lambda=U_{\infty}/f$, is assumed to be small. The vortex sheet, shed tangentially from the trailing edge (by the Kutta condition; see \citet{Jones2003separatedflowinviscid}), is assumed to lie along the $x$-axis downstream of the wing trailing edge (as in the schematic diagram in figure~{\ref{fig:LeaderWake}). We define the vortex sheet strength $\gamma(x,t)$ at point $(x,0)$ as 
\begin{equation}
\gamma(x,t)=u(x,+0,t) - u(x,-0,t).
\end{equation}
We consider the vortex sheet in the wing frame; the sheet defined on $x\ge 1$ downstream the wing trailing edge ($x=1$).
According to Eqs.~(\ref{Eq:linearEuler}) and (\ref{Eq:PhiJump}), the vortex sheet strength~$\gamma$ satisfies
\begin{equation}
\partial_t\gamma + U_{\infty}\partial_x\gamma = 0,\quad x\ge 1,
\end{equation} 
which shows that $\gamma$ is dependent only on a single variable $(x-U_{\infty} t)$, and
\begin{equation}\label{Eq:lagrangian gamma}
\gamma(x,t) = \gamma \left(1,t-\frac{(x-1)}{U_{\infty}}\right).
\end{equation}
The vortex sheet strength at the trailing edge is calculated by \citet{Wu1961Swimmingwavingplate}, as:
\begin{equation}\label{eq:WuGammaFinal}
\gamma(1,t)=\frac{4\pi^{2}A}{G(\sigma)}\sin\left(2\pi\left(t-\frac{\sigma}{2\pi}-g(\sigma)\right)\right),\quad \sigma=2\pi/U_{\infty},
\end{equation}
\begin{equation}\label{Eq:calG}
\mathcal{G}(\sigma)=K_{0}(i\sigma)+K_{1}(i\sigma)=G(\sigma)e^{2\pi ig(\sigma)},
\end{equation}
where $G(\sigma)$ and $g(\sigma)$ are the modulus and angle of $\mathcal{G}(\sigma)$, respectively.
Substituting Eq.~(\ref{eq:WuGammaFinal}) in Eq.~(\ref{Eq:lagrangian gamma}), the vortex sheet strength is then provided as
\begin{equation}\label{Eq:gamma_x_t}
\gamma(x,t)=\frac{4\pi^{2}A}{G(\sigma)}\sin\left(2\pi\left(t-\frac{\sigma x}{2\pi}-g(\sigma)\right)\right),\quad \sigma=2\pi/U_{\infty},
\end{equation}
where $x\ge 1$ denotes the downstream of the wing trailing edge.

\begin{figure}
\centering
\includegraphics[width=\textwidth]{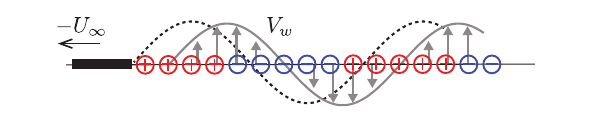}
\caption{\label{fig:LeaderWake}{The vortex sheet generated by a single flapping wing with swimming velocity $-U_{\infty}$ (in the negative $x$-direction) is assumed to lie along the $x$-axis downstream of the wing. It has $\delta_1$ phase-lag from the wing trailing edge trajectory (dashed line). The wake induced by the flat vortex sheet can be approximated by a steady stationary wave~$V_w$, which has $\delta_2$ phase-lag from the wing trailing edge trajectory. See Eq.~(\ref{Eq:gammatilde}) and Eq.~(\ref{Eq:wave by leader}). 
The red $\varoplus$ denote positive vorticity and the blue $\varominus$ denote negative vorticity.}}
\end{figure}

Now we change back to the flow frame, where the wing swims with speed $-U_{\infty}$ towards the negative $x$-direction in a quiescent environment. 
The flow frame variables $(\tilde{x},\tilde{y},\tilde{t})$ and the wing frame variables $(x,y,t)$ are related through 
\begin{equation}\label{Eq:ChangeOfFrame}
(\tilde{x},\tilde{y},\tilde{t})=(x-U_{\infty}t,y,t).
\end{equation}
The position of the wing trailing edge $(\tilde{x}_b,\tilde{y}_b)$ in the flow frame can be parameterized in~$\tilde{t}$, as 
\begin{equation}
\tilde{x}_b = -U_{\infty} \tilde{t}+1,\quad \tilde{y}_b = A\cos\left(2\pi \tilde{t}\right),
\end{equation}
from which the trajectory of the trailing edge is thus provided as
\begin{equation}\label{Eq:ytilde}
\tilde{y}_b(\tilde{x})=A\cos\left(2\pi\frac{(1-\tilde{x})}{U_{\infty}}\right)=A\cos\left(2\pi\frac{(\tilde{x}-1)}{\lambda}\right).
\end{equation} 
Here the wavelength of the motion is $\lambda=U_{\infty}=2\pi/\sigma$ using dimensionless variables.
Substituting Eq.~(\ref{Eq:ChangeOfFrame}) in Eq.~(\ref{Eq:gamma_x_t}), we obtain that the vortex sheet strength in the flow frame is given by
\begin{equation}\label{Eq:gammatilde}
\gamma(\tilde{x})=-\frac{4\pi^{2}A}{G(\sigma)}\sin\left(2\pi\left(\frac{\tilde{x}}{\lambda}+g(\sigma)\right)\right),\quad \tilde{x}\ge \tilde{x}_b,
\end{equation}
where $\tilde{x}_b=-U_{\infty} \tilde{t}+1$ denotes the location of the wing trailing edge.

We note that the vortex sheet strength Eq.~(\ref{Eq:gammatilde}) has a phase-lag of $\delta_1$ from the trailing edge trajectory Eq.~(\ref{Eq:ytilde}), where
\begin{equation}\label{Eq:delta1}
\delta_1(\sigma) = -1/4-g(\sigma)-\sigma/(2\pi).
\end{equation}
In the limiting case of $\sigma\rightarrow 0$, we obtain 
\begin{equation}\label{Eq:G g limit}
G(\sigma)\rightarrow \infty,\;  g(\sigma)\rightarrow -1/4,
\end{equation}
using Eqs.~\eqref{Eq:K0aym}--\eqref{Eq:K1aym} and (\ref{Eq:calG}). Substituting Eq.~(\ref{Eq:G g limit}) in Eqs.~\eqref{Eq:gammatilde}--\eqref{Eq:delta1}, it then follows that the magnitude of the vortex sheet strength $\gamma_0 = 4\pi^2 A/G(\sigma)\rightarrow 0$, and the phase lag $\delta_1(\sigma)\rightarrow 0$ as $\sigma\rightarrow 0$. These limits indicate a weak vortex sheet wake when the wavelength $\lambda$, or the swimming speed $U_{\infty}$ is large. Moreover, the vorticity  shed by the trailing edge is nearly in-phase with the motion of the trailing edge, in the regime of large $\lambda$ or $U_{\infty}$. This agrees 
with experimental flow visualizations in \citet{Ramananarivo2016Flowinteractionslead}, which shows that vortex cores are roughly located at the peak of the trailing edge trajectory.
The limit of $\sigma\rightarrow+\infty$ is of less interest as the steady speed of free swimming $U_{\infty}$ does not achieve small values for the parameter values of our interest. For the range $U_{\infty}\in[2,\infty)$ (i.e. $\sigma \in (0,\pi]$), which covers the range of $U_{\infty}$ in our simulations, we compute numerical ranges $0 < 1/G(\sigma) \apprle 0.705$, and $-0.12 \apprle \delta_1(\sigma) < 0$ which indicates that the phase-lag $\delta_1$ between the trailing edge trajectory and the vorticity is always less than a quarter.

The fluid velocity induced by the vortex sheet Eq.~(\ref{Eq:gammatilde}) can be calculated through the Biot-Savart law. To calculate the flow velocity on the $x$-axis downstream the wing (or the mean velocity at the vortex sheet), we ignore the circulation on the wing body, and approximate the flow field using that induced by the vortex sheet, provided as 
\begin{equation}\label{Eq:linearBiot}
u(\tilde{x},0,t)-iv(\tilde{x},0,t)=\frac{1}{2\pi i}\Xint-_{\tilde{x}_b}^{\infty}\frac{\gamma(\tilde{x}')}{\tilde{x}-\tilde{x}'}\,\d\tilde{x}', \quad \tilde{x}\ge \tilde{x}_b,
\end{equation}
where $\tilde{x}_b$ denotes the location of the wing trailing edge. Note that $u(\tilde{x},0,t)=0$, as the right hand side of Eq.~(\ref{Eq:linearBiot}) is purely imaginary. While the horizontal velocity on the $x$-axis, downstream the wing, is zero, the vertical velocity is given by
\begin{eqnarray}
v(\tilde{x},0,t) & = & \frac{1}{2\pi}\Xint-_{\tilde{x}_b}^{+\infty}\frac{\gamma(\tilde{x}')}{\tilde{x}-\tilde{x}'}\,\d\tilde{x}'\nonumber\\
& = & -\frac{2\pi A}{G(\sigma)}\Xint-_{\tilde{x}_b}^{+\infty}\frac{\sin\left(2\pi(\tilde{x}/\lambda+g(\sigma))\right)}{\tilde{x}-\tilde{x}'}\,\d\tilde{x}'\nonumber\\
 & = & \frac{2\pi^2 A}{G(\sigma)}\left\{ \cos\left(2\pi\left(\frac{\tilde{x}}{\lambda}+g(\sigma)\right)\right)+J(\tilde{x},\tilde{x}_b)\right\},\label{eq:wave}
\end{eqnarray}
\begin{eqnarray}\label{Eq:J}
J(\tilde{x},\tilde{x}_b)  & = & \frac{1}{\pi}\left[\left(\Si(2\pi(\tilde{x}-\tilde{x}_b)/\lambda) -\tfrac{1}{2}\pi\right)\cos\left(2\pi(\tilde{x}/\lambda+g(\sigma))\right)\right.\nonumber \\
& &\left. -\Ci(2\pi(\tilde{x}-\tilde{x}_b)/\lambda)\sin\left(2\pi(\tilde{x}/\lambda+g(\sigma))\right)\right], \quad \tilde{x}\ge \tilde{x}_b,
\end{eqnarray}
where $\Si(x)=\int_0^x\frac{\sin t}{t}\,\d t$ and $\Ci(x)=-\int_x^{+\infty}\frac{\cos t}{t}\,\d t$ are the sine integral and cosine integral of real arguments, respectively. 

Note that Eq.~(\ref{Eq:J}) can be bounded by its magnitude function $I(\tilde{x}-\tilde{x}_b)$, where
\begin{equation}
I(x) =  \sqrt{\left(\Si(2\pi x)-\tfrac{1}{2}\pi\right)^{2}+\Ci^{2}(2\pi x)}.
\end{equation}
It is easy to show that $I(x)$ is a monotonically decreasing function. For location $\tilde{x}$ that is at least one wavelength away from the wing trailing edge $\tilde{x}_b$, i.e. $(\tilde{x}-\tilde{x}_b)/\lambda \ge 1$, Eq.~(\ref{Eq:J}) is bounded numerically, as follows:
\begin{equation}
|J(\tilde{x},\tilde{x}_b)| \le \frac{1}{\pi} I((\tilde{x}-\tilde{x}_b)/\lambda)) \le I(1) \approx 0.05.
\end{equation}
In Eq.~(\ref{eq:wave}), the magnitude of $J(\tilde{x},\tilde{x}_b)$ is small compared to  $\cos\left(2\pi(\tilde{x}/\lambda+g)\right)$, the other term in the bracket. Therefore, $J(\tilde{x},\tilde{x}_b)$ can be neglected and Eq.~(\ref{eq:wave}) can be approximated in a simple wave form (figure~\ref{fig:LeaderWake}), as
\begin{equation}\label{Eq:wave by leader}
v(\tilde{x},0,t) \approx V_{w}(\tilde{x}) = \frac{2\pi^2 A}{G(\sigma)} \cos\left(2\pi\left(\frac{\tilde{x}}{\lambda}+g(\sigma)\right)\right).
\end{equation}
Moreover, for large $x$, the asymptotic expansion 
$I(x)={1}/{(2\pi^2 x)} + O(x^{-3})$ is obtained 
from expansions of $\Si(x)$ and $\Ci(x)$. It then follows that $|J(\tilde{x},\tilde{x}_b)|\rightarrow 0$ for ${(\tilde{x}-\tilde{x}_b)/\lambda\rightarrow+\infty}$, which indicates that the approximation $V_{w}(\tilde{x})$ in Eq.~(\ref{Eq:wave by leader}) is closer to $v(\tilde{x},0,t)$ for $\tilde{x}$ further away the wing trailing edge $\tilde{x}_b$.
Note that the wave $V_{w}(\tilde{x})$ (Eq.~(\ref{Eq:wave by leader})) is a quarter lag in phase from the vorticity distribution $\gamma(\tilde{x})$ (Eq. (\ref{Eq:gammatilde})). Therefore, the phase difference between the wave $V_{w}(\tilde{x})$ and the trailing edge trajectory $\tilde{y}_b(\tilde{x})$ (Eq.~(\ref{Eq:ytilde})) is then
\begin{equation}\label{Eq:delta2}
\delta_2(\sigma) =\delta_1(\sigma)+1/4 =  -g(\sigma)-\frac{\sigma}{2\pi},
\end{equation}
and it has the limit $\delta_2(\sigma)\rightarrow 1/4$ as $\sigma\rightarrow 0$.

The steady stationary wave $V_{w}(\tilde{x})$ in Eq.~(\ref{Eq:wave by leader}) provides a simple model of the flow at the centerline of a reverse von K\'arm\'an vortex street, in the limit of small Strouhal number $St=Af/U_{\infty}.$ 
For the parameters used in both the schooling experiments and our simulations, the flapping velocity $V_f=2\pi Af$ is
small compared to the swimming velocity $U_{\infty}$ so that the Strouhal number is small. In the experiments of \citet{Ramananarivo2016Flowinteractionslead}, $\St=0.1$ to $0.2$, and in our simulations $\St=0.005$ to $0.1$.
From experimental flow visualization of a single flapping wing \citep{Becker2015Hydrodynamicschoolingflapping,Ramananarivo2016Flowinteractionslead} and our simulations (see the wake of the leader wing in figure~\ref{fig:potential_wake}), we find that the backward jet flow at the centerline in the reverse von K\'arm\'an street is weak, as the Strouhal number is small. Flow visualization shows alternating upward and downward flow jets along the centerline of the wake generated by a single flapping wing. This allows us to use a waveform (Eq.~(\ref{Eq:wave by leader})) to model the flow at the centerline of the wake produced by the leader wing in the school, and subsequently study its interaction with the follower wing.

\subsubsection{Single flapping wing interacting with steady wave}
Now we consider a single flapping wing interacting with a wavy stream in the background (figure~\ref{fig:FollowerInteraction}). It serves as a model for the follower swimmer in the ``school-of-two," that is surrounded by the wake produced by the leader wing. In the frame of the wing, where the wing spans from $-1 \le x \le 1$, we assume the wing is immersed in a wavy stream with velocity $(U_{\infty},V_{sw}(x,t)).$ The $x$-component of this background flow is a uniform stream $U_{\infty},$ and the $y$-component $V_{sw}$ is a traveling wave of the same velocity $U_{\infty}$ and the frequency is $f$, the same frequency as the flapping wing. We denote the wavelength by $\lambda=U_{\infty}/f$. Using dimensionless variables, where $f=1$, the traveling wave is given~by
\begin{equation}\label{Eq:wavyWing}
V_{sw}(x,t)=V_{0}\sin(2\pi(x/\lambda-t)),\quad \lambda=U_{\infty}/f,
\end{equation}
where $V_0$ is the amplitude of the wave,
and the heaving motion of the wing has the form
\begin{equation}\label{Eq:hxtwave}
h(x,t)=A\cos(2\pi(t+\theta)),\quad -1\le x \le 1,
\end{equation}
where $\theta$ is a phase-shift in time, and the flapping frequency ($f=1$) is the same as the traveling wave. 

The linearized problem of a rigid wing flapping in a wavy stream is solved by \citet{Wu1975Extractionflowenergy}. There, the wing maintains a superposition motion of pitching and heaving, with the flapping frequency required to be the same as the frequency of the wavy stream. In this work, we adapt the calculation of \citet{Wu1975Extractionflowenergy} to our case of a purely heaving wing (Eq.~(\ref{Eq:hxtwave})), flapping in a wavy stream $(U_{\infty},V_{sw})$ with the wave speed the same as the stream speed.

\begin{figure}
\centering
\includegraphics[width=.65\textwidth]{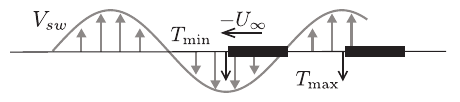}
\caption{\label{fig:FollowerInteraction}{A single flapping wing swimming with velocity $-U_{\infty}$ (negative $x$-direction) interacts with a stationary wave $V_{sw}$ in the background. When the leading edge velocity of the wing is in-phase with the wave, the effective wing flapping is reduced and the wing thrust reaches a minimum value $T_{\min}$. When the wing leading edge is out-of-phase with the wave, the effective flapping speed is increased and the thrust reaches a maximum value~$T_{\max}$.}}
\end{figure}

We assume the heaving amplitude $A$ and the amplitude of the wave $V_0$ are small, and denote $\epsilon=V_{0}/(2\pi A)=O(1)$ (using dimensionless variables; $\epsilon=V_{0}/(2\pi Af)$ if using dimensional variables) the ratio of the wave velocity to the flapping velocity. Similar to the linearization of a single wing in a quiescent background (see Section~\ref{sec: singleQues}), we denote the flow velocity by $\mathbf{q} = (U_{\infty}+u,V_{sw}+v).$ Assuming $U_{\infty} \gg u$ and $U_{\infty} \gg v$, the linearized 2D Euler equation Eq.~(\ref{Eq:linearEuler}) has the following kinematic boundary condition,
\begin{equation}
v_{1}(x,y,t)\mid_{y=\pm0}=\partial_t h(x,t)-V_{sw}(x,t),\quad -1\le x\le 1.
\end{equation}
Using the same assumptions Eqs.~(\ref{Eq:PhiJump})--(\ref{Eq:fardecay}) in Section~\ref{sec: singleQues}, the acceleration potential $f(z,t)$ (defined in Eq.~(\ref{Eq:acceleration potential})) is solved by \citet{Wu1975Extractionflowenergy} through Fourier
series analysis and conformal mapping techniques, a process similar to that in \citet{Wu1961Swimmingwavingplate}. The stroke-averaged thrust on the wing is provided by \citet{Wu1975Extractionflowenergy}, as
\begin{equation} \label{Eq:ThrustInWave}
\langle T \rangle = -4\pi^3 A^{2}\left|C(\sigma)\right|^{2}\left[ 1+\epsilon^{2}P^{2}(\sigma)+2\epsilon P(\sigma)\sin(2\pi\left(\theta-p(\sigma)\right))\right],
\end{equation}
\begin{equation}\label{Eq:calP}
\mathcal{P}(\sigma) =\frac{\overline{K_{1}(i\sigma)}K_{0}(i\sigma)+\overline{K_{0}(i\sigma)}K_{1}(i\sigma)}{\pi K_{1}(i\sigma)}= P(\sigma)e^{2\pi ip(\sigma)},
\end{equation}
where $\epsilon=V_{0}/(2\pi A)$, $\sigma=2\pi/U_{\infty},$ $P(\sigma)$ and $p(\sigma)$ are the modulus and angle of $\mathcal{P}(\sigma)$, respectively. Note that the thrust Eq.~(\ref{Eq:ThrustInWave}) has a simple sinusoidal form of three variables $\theta$, $\epsilon$ and $\sigma$, where $\theta$ denotes the phase difference between the wave and the wing flapping, $\epsilon$ denotes the ratio of the flapping speed and the wave speed, and $\sigma=2\pi/U_{\infty}=2\pi/\lambda$ ($\sigma=\pi cf/U_{\infty}=\pi c/\lambda$ using dimensional quantities) represents the wavelength $\lambda$ of the wave, or stream speed $U_{\infty}$.

In the flow frame, where the $x$-component of the background flow is zero and the wing translates with velocity $-U_{\infty}$ in the negative $x$-direction, by substituting Eq.~(\ref{Eq:ChangeOfFrame}) in Eq.~(\ref{Eq:wavyWing}), the background wave Eq.~(\ref{Eq:wavyWing}) now becomes a steady vertical wave,
\begin{equation}\label{eq:StationaryWave}
\tilde{V}_{sw}(\tilde{x})=V_{0}\sin(2\pi \tilde{x}/\lambda).
\end{equation}
The leading edge of the wing, parameterized in $\tilde{t}$ as,
\begin{equation}
\tilde{x} = -U_{\infty} \tilde{t}-1,\quad \tilde{y} = A\cos(2\pi (\tilde{t}+\theta)),
\end{equation}
follows  the trajectory 
\begin{equation}\label{Eq:ytilde2}
\tilde{y}(\tilde{x})=A\cos(2\pi\left((\tilde{x}+1)/\lambda-\theta\right)).
\end{equation} 
Using Eqs.~(\ref{Eq:ytilde2}) and (\ref{eq:StationaryWave}), we obtain the phase lag of the wing's leading edge $\tilde{y}(\tilde{x})$ to the vertical wave $\tilde{V}_{sw}(\tilde{x})$, that is,
\begin{equation}\label{Eq:tildetheta}
\tilde{\theta} =  \theta - \sigma/(2\pi) - 1/4,
\end{equation}
where $\sigma = 2\pi/\lambda,$
and the thrust on the wing (Eq.~(\ref{Eq:ThrustInWave})) can be expressed in terms of $\tilde{\theta}$ as,
\begin{equation} \label{Eq:wavy thrust}
\langle T \rangle = -4\pi^3 A^{2}\left|C(\sigma)\right|^{2}\left[ 1+\epsilon^{2}P^{2}(\sigma)+2\epsilon P(\sigma)\cos\left(2\pi\left(\tilde{\theta}-\delta_3(\sigma)\right)\right)\right],
\end{equation}
\begin{equation}\label{Eq:delta3}
\delta_3(\sigma) =  p(\sigma)-\sigma/(2\pi).
\end{equation}

It is shown by Eq.~(\ref{Eq:wavy thrust}) that the extrema of~$\langle T \rangle$ are determined by the phase difference between the wing and the wave $\tilde{\theta}$, which is of particular interest. Extreme values of thrust~$\langle T \rangle$ are
\begin{eqnarray}
\langle T \rangle_{\max}  & = & -4\pi^3 A^{2}\left|C(\sigma)\right|^{2}\left[ 1+\epsilon P(\sigma)\right]^2, \quad \text{when} \quad \tilde{\theta}=\delta_3 + k,\label{Eq:Tmax wave}\\
\langle T \rangle_{\min} & = & -4\pi^3 A^{2}\left|C(\sigma)\right|^{2}\left[ 1-\epsilon P(\sigma)\right]^2, \quad \text{when} \quad \tilde{\theta}=\delta_3 + k + 1/2,\label{Eq:Tmin wave}
\end{eqnarray}
where $k\in \mathbb{Z}^+$. For $\lambda  \in [2,\infty)$, i.e. $\sigma = 2\pi/\lambda \in (0,\pi]$, which is the main range of interest, the numerical value is approximately $0.143 \apprle \delta_3 \le 1/4$, which is about a quarter. Moreover, in the limit of $\sigma \rightarrow 0$, we obtain $P(\sigma)\rightarrow 1$ and $p(\sigma)\rightarrow 1/4$ by substituting asymptotic expansions of the modified Bessel functions Eqs.~(\ref{Eq:K0aym})--(\ref{Eq:K1aym}) in Eq.~(\ref{Eq:calP}), and thus ${\delta_3(\sigma) \rightarrow 1/4}$. 
These values indicate that the maximum thrust is achieved when the wing leading edge trajectory Eq.~(\ref{Eq:ytilde2}) has phase lag $\tilde{\theta} \approx k+1/4$ to the stationary wave Eq.~(\ref{eq:StationaryWave}),
and the minimum thrust is achieved when the leading edge trajectory has about $\tilde{\theta}\approx k+3/4$ phase lag to the wave. 

This could be explained by looking at the flapping velocity of the wing leading edge, which can be parameterized as
\begin{equation}
\tilde{x} = -U_{\infty} \tilde{t}-1,\quad \dot{\tilde{y}} = -2\pi A\sin(2\pi (\tilde{t}+\theta)),
\end{equation}
by Eq.~(\ref{Eq:ChangeOfFrame}), and be expressed in $\tilde{x}$ as
\begin{equation}\label{Eq:dotytilde2}
\dot{\tilde{y}}(\tilde{x})=2\pi A\sin\left(2\pi\left(\tilde{x}/\lambda-\tilde{\theta}-1/4 \right)\right).
\end{equation}
We define an ``effective flapping" velocity at the wing leading edge,
\begin{equation}
V_{\mathrm{eff}}(\tilde{x})= \dot{\tilde{y}}(\tilde{x})-\tilde{V}_{sw}(\tilde{x})=2\pi A\sin\left(2\pi\left(\tilde{x}/\lambda-\tilde{\theta}-1/4 \right)\right)-V_{0}\sin(2\pi \tilde{x}/\lambda),
\end{equation}
which is the relative velocity of the wing flapping Eq.~(\ref{Eq:dotytilde2}) and the background wave Eq.~(\ref{eq:StationaryWave}). Note that the maximum thrust Eq.~(\ref{Eq:Tmax wave}) is achieved when $\tilde{\theta} \approx k+1/4$, at which time the flapping velocity of the leading edge Eq.~(\ref{Eq:dotytilde2}) is {\it out-of-phase} with the wave, and the effective flapping has a maximum amplitude 
\begin{equation}
V_{\mathrm{eff},\max}(\tilde{x}) = (2\pi A + V_0)\sin\left(2\pi\tilde{x}/\lambda \right),
\end{equation}
as shown in the schematic diagram in figure~\ref{fig:FollowerInteraction}.
Similarly, the minimum thrust corresponds to the heaving velocity at the leading edge {\it in-phase} with the wave, and the amplitude of effective flapping achieves minimum
\begin{equation}
V_{\mathrm{eff},\min}(\tilde{x})=(2\pi A-V_0)\sin\left(2\pi\tilde{x}/\lambda \right).
\end{equation}
These results are not surprising. As the thrust $\langle T \rangle$ consists only of the suction force at the leading edge of the wing, 
when the background wave is out-of-phase with the flapping velocity of the wing leading edge, the effective flapping is increased and the wing thrust is thus increased and reaches a maximum. Similarly when the wave is in-phase with the flapping of the wing leading edge, the effective flapping is decreased and the thrust reaches a minimum.

\subsection{Linearized theory adapted to two wings in tandem}
\begin{figure}
\centering
\includegraphics[width=\textwidth]{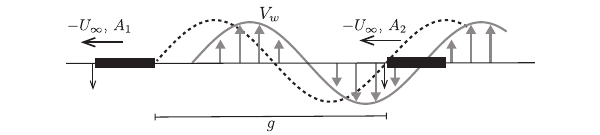}\caption{\label{fig:SchoolingWu}{Adapted linearized theory to schooling of two wings in tandem. The wings flap with amplitude $A_1$ and $A_2$, respectively, and swim with the same velocity $-U_{\infty}$ in the negative $x$-direction. The leader generates a spatial wave $V_w$ which produces a hydrodynamic potential on the follower. The dashed line denotes the trailing edge trajectory of the leader wing. The ``tail-head" separation between the wings is denoted by $g$. }}
\end{figure}
Now we adapt the linearized calculation on individual wings to the school of two wings swimming in tandem. The two wings are assumed to have the same swimming velocity $-U_{\infty}$ in the negative $x$-direction, while the leader wing has flapping amplitude~$A_1$ and the follower has amplitude $A_2$ (see schematics in figure~\ref{fig:SchoolingWu}). To adapt the linearized theory to schooling, we approximate the leader wing using calculations of an isolated single wing, assuming the leader  maintains a constant motion and neglecting the hydrodynamic perturbation induced by the follower to the leader. This is motivated by previous simulation results (Section~\ref{sec:simulation}) that the speed of the leader wing stays always close to the speed of single wing for the follower not following closely (schooling number $S$ not too small), regardless of the location and flapping amplitude of the follower wing (see figures~\ref{fig:school1}(b) and \ref{fig:potential}(a)). It is also supported by the experiments of \citet{Ramananarivo2016Flowinteractionslead}, which show that the schooling speed is always close to the single wing speed, with a difference of about $10\%$ for most cases. The wake generated by the leader wing is approximated by a spatial wave of amplitude $V_0=2\pi^2A_1/G(\sigma)$ and phase-lag $\delta_2(\sigma)$ to the wing trailing edge, where $\sigma = 2\pi/U_{\infty}$ (Eqs.~(\ref{Eq:wave by leader}) and (\ref{Eq:delta2})). We calculate the hydrodynamic potential that the leader induces on the follower by considering the interaction of the follower with this stationary wave, neglecting perturbations due to the body of the leader wing.

\subsubsection{Hydrodynamic thrust of the follower wing}
We denote the distance between the follower's head and the leader's tail by $g$ (as in figure~\ref{fig:SchoolingWu}). The schooling number defined as $S=g/\lambda$ ($\lambda=U_{\infty}/f$) denotes the phase-lag of the follower's head to leader's tail, following which 
the phase-lag of the follower's leading edge to the stationary wave generated by the leader is
\begin{equation}\label{Eq:tilde theta s}
\tilde{\theta} = S - \delta_2(\sigma).
\end{equation}
Substituting Eq.~(\ref{Eq:tilde theta s}) in Eq.~(\ref{Eq:wavy thrust}), and by Eqs.~(\ref{Eq:calG}), (\ref{Eq:delta2}), (\ref{Eq:calP}), and (\ref{Eq:delta3}), we express the thrust on the follower in schooling number $S$, as,
\begin{equation} \label{Eq:follower thrust}
\langle T_2 \rangle  = \langle T_1 \rangle \left[ \varepsilon^{2}+Q^{2}(\sigma)+2\varepsilon Q(\sigma)\cos\left(2\pi\left(S - \delta_4(\sigma)\right)\right)\right],\quad \varepsilon=A_2/A_1,
\end{equation}
\begin{equation}\label{Eq:delta4}
\delta_4(\sigma) = \delta_2(\sigma)+\delta_3(\sigma)=q(\sigma)-\sigma/\pi,
\end{equation}
where $\langle T_1 \rangle = -4\pi^3 A_1^{2}\left|C(\sigma)\right|^{2}$ is the thrust of the leader wing (Eq.~(\ref{Eq:T_single})), $Q(\sigma)$ and $q(\sigma)$ are the modulus and angle of 
\begin{equation}\label{Eq:calQ}
\mathcal{Q}(\sigma) = \frac{\pi\mathcal{P}(\sigma)}{\mathcal{G}(\sigma)} =\frac{\overline{K_{1}(i\sigma)}K_{0}(i\sigma)+\overline{K_{0}(i\sigma)}K_{1}(i\sigma)}{K_{1}(i\sigma)\left[K_{0}(i\sigma)+K_{1}(i\sigma)\right]}  =Q(\sigma)e^{2\pi iq(\sigma)},
\end{equation}
respectively.
In figure~\ref{fig:LinearPotential}(a), the follower's thrust (black solid curve) calculated from Eq.~(\ref{Eq:follower thrust}) is compared with the thrust computed by the vortex sheet simulations (gray solid curve) in Section~\ref{sec:simulation}, in which the thrust is measured on a ``ghost follower" maintaining a fixed separation to the leader.
Here the two wings have the same amplitude $A=0.2$, the drag coefficient is fixed at $C_f=0.02$, and  $\sigma$ in Eq.~(\ref{Eq:follower thrust}) is calculated from Eq.~(\ref{Eq:linear balance}).

\begin{figure}
\centering\includegraphics[width=\textwidth]{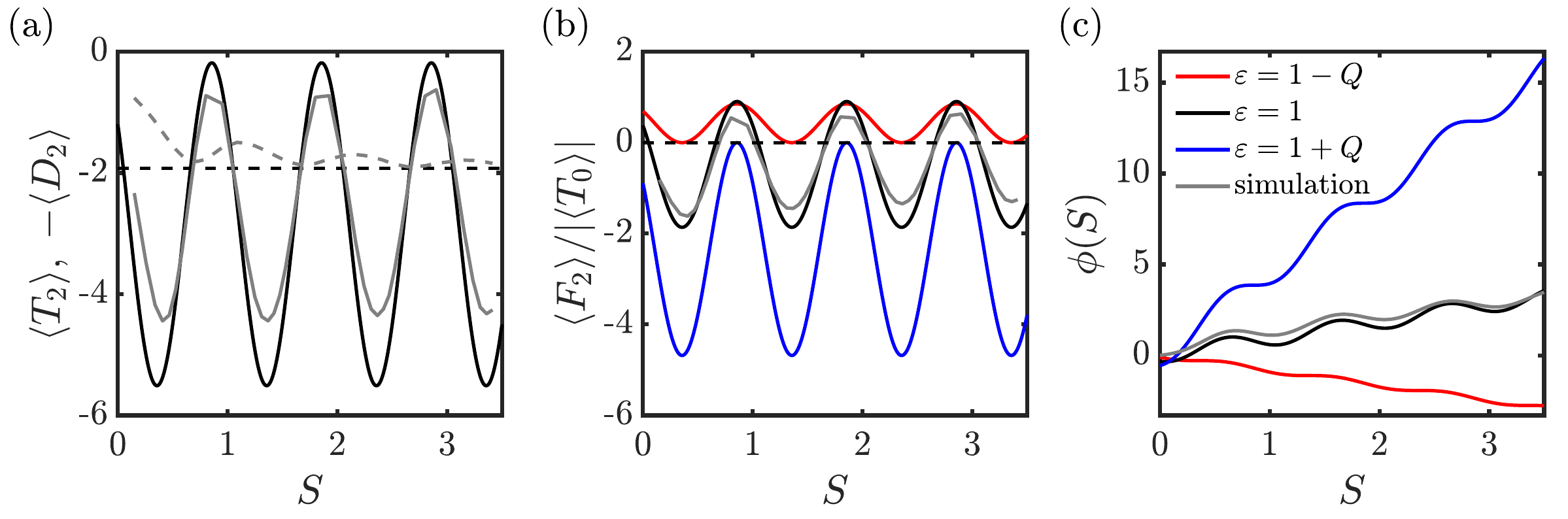}
\caption{\label{fig:LinearPotential}{Comparisons between the hydrodynamic forces and potential on the follower wing calculated from linearized theory (black) and from simulations (gray). (a) Hydrodynamic thrust $\langle T_2\rangle$ (solid lines) and drag $-\langle D_2\rangle$ (dashed lines) as functions of schooling number $S$, where the two wings have the same flapping amplitude $A=0.2$. The forces here are dimensionless. (b) Total hydrodynamic force on the follower $\langle F_2\rangle$ normalized by the single wing thrust~$|\langle T_0\rangle |$. (c) Hydrodynamic potential on the follower. In (b)--(c), the leader's amplitude is fixed at $A_1=0.2$, while the follower's amplitude $A_2$ is varied. Here $\varepsilon=A_2/A_1=1-Q$ (red), $1$ (black and gray), and $1+Q$ (blue). In all cases the drag coefficient is $C_f=0.02$.}}
\end{figure}

Note that the follower's thrust Eq.~(\ref{Eq:follower thrust}) is a sinusoidal function in schooling number~$S$, i.e. it depends on the phase difference between the two wings swimming. The coefficients of this sinusoidal function depend on the amplitude ratio of the wings $\varepsilon=A_2/A_1$ and the wavelength of the motion $\lambda/c=2 \pi/\sigma$ (dimensionless). For $\varepsilon=0$, Eq.~\eqref{Eq:follower thrust} simplifies to
\begin{equation}
\langle T_2 \rangle|_{\varepsilon=0} = \langle T_1 \rangle Q^2(\sigma)=-4\pi^3 \left|C(\sigma)\right|^{2}\left( Q(\sigma)A_1\right)^2.
\end{equation}
This implies that when the follower wing is a non-flapping ``dead wing," if it somehow still maintains a swimming speed $U_{\infty}$, it can generate a non-zero constant thrust due to the interaction with the wavy wake shed by the leader. This ``dead-follower" thrust $\langle T_2 \rangle|_{\varepsilon=0}\to 0$ at the long wavelength limit $\sigma\rightarrow 0$, as $Q(\sigma)\rightarrow 0$ (Eqs.~(\ref{Eq:K0aym})--(\ref{Eq:K1aym}) and~\eqref{Eq:calQ}). 

The follower's thrust Eq.~(\ref{Eq:follower thrust}) achieves extreme values
\begin{eqnarray}
\langle T_2 \rangle_{\max}  & = & \langle T_1 \rangle \left[ \varepsilon+Q(\sigma)\right]^2=-4\pi^3 \left|C(\sigma)\right|^{2}\left[ A_2+Q(\sigma)A_1\right]^2,\; \text{when} \; S=\delta_4+k,\nonumber \\
\langle T_2 \rangle_{\min}  & = & \langle T_1 \rangle \left[ \varepsilon-Q(\sigma)\right]^2=-4\pi^3 \left|C(\sigma)\right|^{2}\left[ A_2-Q(\sigma)A_1\right]^2,\; \text{when} \; S=\delta_4+k+1/2, \nonumber
\end{eqnarray}
where $k\in\mathbb{Z}^+$. In the limiting case of $\sigma\rightarrow 0$, we have $q(\sigma)\rightarrow 1/2$ and thus $\delta_4\rightarrow 1/2$ (Eq.~(\ref{Eq:delta4})). This implies that at the long wavelength limit, $\lambda/c\rightarrow \infty$, the maximum thrust $\langle T_2 \rangle_{\max}$ is achieved when the schooling number $S\approx 1/2+k$ and minimum thrust $\langle T_2 \rangle_{\max}$ occurs at $S\approx k$. Moreover, we note that the perturbation due to the schooling is weak as $Q(\sigma)\rightarrow 0$ when $\sigma\rightarrow 0$. When $\sigma=0$, the follower's thrust $\langle T_2 \rangle = -4\pi^3 A_2^2\left|C(\sigma)\right|^{2}$ is the thrust of a single wing maintaining the follower's flapping amplitude $A_2$.

\subsubsection{Hydrodynamic potential, schooling numbers, and ``keep-up" schooling condition}

We approximate the free stream velocity on wing surfaces, $U_{\infty}+u$ in Eq.~(\ref{Eq:linearEuler}), by the wing swimming speed $U_{\infty}$, ignoring the small perturbation in horizontal velocity $u$, as in the linearized calculation. The skin-friction $D_k(|U_{\infty}|),\ k=1,2$ is then a function of~$U_{\infty}$ only. As the two wings swim at the same speed, they experience the same amount of drag $D_1=D_2$. Note that  the stroke-averaged thrust  and drag are always balanced on the leader wing, $\langle T_{1} \rangle +D_1=0$ (Eq.~(\ref{Eq:linear balance})). It then follows that ${0=\langle T_{1} \rangle +D_1=\langle T_{1} \rangle +D_2}$. Using this relation and Eq.~(\ref{Eq:follower thrust}), we obtain the following expression for the total hydrodynamic force on the follower,
\begin{equation}\label{Eq: F_2}
\langle F_2 \rangle  = \langle T_2 \rangle+ D_2 =\langle T_1 \rangle\left[ \varepsilon^{2}+Q^{2}(\sigma)-1+2\varepsilon Q(\sigma)\cos\left(2\pi\left(S - \delta_4(\sigma)\right)\right)\right].
\end{equation}
In figure~\ref{fig:LinearPotential}(a), the approximation of the follower's drag $D_2=-\langle T_{1} \rangle$, calculated by Eq.~(\ref{Eq:T_single}) (black dashed curve), is compared with the drag computed from the simulations (gray dashed curve), which shows the theoretical approximation is close to the simulation when the schooling number $S$ is not too small, i.e. the follower is not too close to the leader. It then follows that the total hydrodynamic force formula (Eq.~(\ref{Eq: F_2})) serves as a good approximation to the simulations, as shown in figure~\ref{fig:LinearPotential}(b).

The hydrodynamic potential $\phi(S)$ is the negative integral of the the hydrodynamic force $\langle F_2 \rangle $, given as
\begin{equation}\label{Eq: phi_linear}
\phi(S)  = -\int \langle F_2 \rangle \,\d S=-\langle T_1 \rangle\left[\left( \varepsilon^{2}+Q^{2}(\sigma)-1\right)S+\frac{\varepsilon Q(\sigma)}{\pi}\sin\left(2\pi\left(S - \delta_4(\sigma)\right)\right)\right]. 
\end{equation}
In figures~\ref{fig:LinearPotential}(b) and (c), the hydrodynamic force calculated from Eq.~(\ref{Eq: F_2}) and hydrodynamic potential from  Eq.~(\ref{Eq: phi_linear}) are shown, and compared with simulations of two wings having the same flapping amplitude (the same simulation data as in figure~\ref{fig:LinearPotential}(b)).
Here~$\sigma$ in Eqs.~(\ref{Eq: F_2})--(\ref{Eq: phi_linear}) is solved from Eq.~(\ref{Eq:linear balance}) given a flapping amplitude $A=0.2$ and a drag coefficient $C_f=0.02$.

The steady schooling numbers are roots of $\langle F_2 \rangle $, or local extrema of $\phi(S)$, which can be solved from Eq.~(\ref{Eq: F_2}):
\begin{equation}\label{Eq:Sk}
S_{k}^{\pm}=\pm\frac{1}{2\pi}\arccos\left(\frac{1-\varepsilon^2-Q^2(\sigma)}{2\varepsilon Q(\sigma)}\right)+q(\sigma)-\frac{\sigma}{\pi} + k,\quad k\in\mathbb{Z}^+.
\end{equation}
Moreover, it can be shown from Eq.~(\ref{Eq: phi_linear}) that $S_{k}^{-}$ are stable schooling numbers (stable critical points) and $S_{k}^{+}$ are unstable (see figure~\ref{fig:LinearSchoolingNumber}). If the two wings have the same flapping amplitude, $\varepsilon = A_2/A_1=1$, Eq.~(\ref{Eq:Sk}) simplifies to
\begin{equation}\label{Eq:Sk_ep1}
S_{k}^{\pm}=\pm\frac{1}{2\pi}\arccos\left(-Q(\sigma)/2\right)+q(\sigma)-\frac{\sigma}{\pi}+ k,\quad k\in\mathbb{Z}^+.
\end{equation}
This is compared in figures~\ref{fig:LinearSchoolingNumber}(a)--(b) with simulations of two wings having the same flapping amplitude, where $\sigma$ is solved from Eq.~(\ref{Eq:linear balance}) given a flapping amplitude $A$ and a drag coefficient $C_f$. At the long wavelength limit $\sigma=\pi c/\lambda\rightarrow 0$, we have $Q(\sigma)\rightarrow 0$ and $q(\sigma)\rightarrow 1/2$, so that the schooling numbers achieve limits
\begin{equation}
S_{k}^{-}\rightarrow k+\frac{1}{4},\quad S_{k}^{+}\rightarrow k+\frac{3}{4},\quad k\in\mathbb{Z}^+.
\end{equation}

\begin{figure}
\centering
\includegraphics[width=\textwidth]{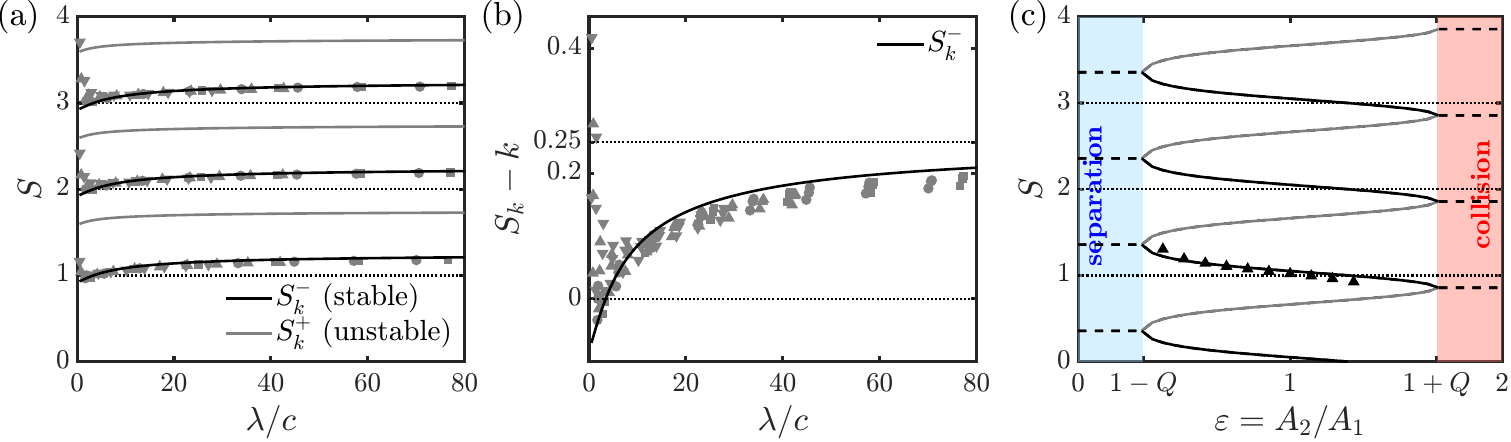}
\caption{\label{fig:LinearSchoolingNumber}{Schooling numbers computed from simulations (symbols) are compared with the linearized theory calculations (lines). Stable schooling numbers $S_k^-$ are black and unstable schooling numbers $S_k^+$ are gray. (a)--(b) Flapping amplitudes are the same on both wings, the amplitude $A$ and drag coefficient $C_f$ are varied. In (b), integers ${k=1,2,3}$ are subtracted from stable schooling numbers $S_k^-$. The limit $S_k^{-}-k\rightarrow 1/4$ as $\lambda/c\rightarrow \infty$. The simulation data is the same as in figure~\ref{fig:school2}. (c) Fix leader's amplitude $A_1=0.2$, and vary follower's amplitude $A_2$. Stable and unstable schooling numbers exist when ${1-Q\le A_2/A_1 \le 1+Q}$. The simulation data is the same as in figure~\ref{fig:bigsmall}(a). The drag coefficient is $C_f=0.02$.} }
\end{figure}

Note that if the two wings have different flapping amplitudes, Eq.~(\ref{Eq:Sk}) indicates a condition for steady schooling states, i.e.,
\begin{equation}
-1\le \frac{1-\varepsilon^2-Q^2(\sigma)}{2\varepsilon Q(\sigma)} \le 1.
\end{equation}
This provides a ``keep-up" condition for the ratio of the follower's amplitude $A_2$ to the leader's amplitude $A_1$:
\begin{equation}\label{Eq:A2A1}
1-Q(\sigma)\le A_2/A_1 \le 1+Q(\sigma).
\end{equation}
As figure~\ref{fig:LinearSchoolingNumber}(c) shows, the schooling number exists only for $1-Q(\sigma)\le \varepsilon \le 1+Q(\sigma)$. Stable and unstable branches merge at $\varepsilon=1\pm Q(\sigma).$ It can also be seen in figures~\ref{fig:LinearPotential}(b)--(c) that for $\varepsilon=1\pm Q(\sigma),$ the hydrodynamic force $\langle F_2 \rangle$ in Eq.~(\ref{Eq: F_2}) has only one root in each period, where the stable and unstable solutions collapse. Beyond the range of Eq.~(\ref{Eq:A2A1}), $\langle F_2 \rangle$ does not intersect with the $x$-axis. When $A_2/A_1 < 1-Q(\sigma)$, the hydrodynamic force on the follower wing is always strictly positive (always in the opposite direction of swimming). In this case, the follower cannot generate enough force maintaining the same swimming speed $U_{\infty}$ as the leader, so that the follower is not able to ``keep-up" in the school and will drift away downstream. It can also be explained by the hydrodynamic potential profile $\phi(S)$ in figure~\ref{fig:LinearPotential}(c), that $\phi(S)$ is a monotonically decreasing function in~$S$, and wherever the follower locates in this potential the schooling number $S$ could not be trapped and will keep increasing. For $A_2/A_1 > 1+Q(\sigma)$, the hydrodynamic force is always strictly negative (always in the swimming direction), therefore the follower will keep accelerating, eventually leading to ``rear-ending" collisions within the school.

\section{Discussion and conclusions}\label{sec:discussion}

We have studied theoretically the hydrodynamics and locomotion dynamics of interacting flapping wings or foils in tandem. Numerical simulations show that the follower wing of a pair, when released to swim freely, tends to re-position itself within the leader's wake flow and lock into one of multiple positions. These emergent stable states correspond to so-called schooling numbers $S=g/\lambda$ near integer values, indicating that the follower takes up a separation distance $g$ that is nearly an integer multiple of the swimming wavelength~$\lambda$. This observation is consistent with previous experiments~\citep{Ramananarivo2016Flowinteractionslead,newbolt2019flow,newbolt2024flow} and vortex-sheet simulations~\citep{heydari2021school,nitsche2024stability}. A systematic analysis of the stable configurations shows that the emergent~$S$ values are a weak function of $\lambda$, i.e. $S_k\approx S(\lambda)+k,\ k\in \mathbb{Z}^+$. We have also mapped out the hydrodynamic potential on the follower wing using two methods: one in which a fixed force is applied and the resulting position is determined, and the other in which a fixed position within the leader's wake is imposed and the resulting force is determined. Both methods show that the interaction potential resembles a tilted washboard: 
a series of local minima or stable wells, corresponding to the observed `schooling' states, superposed on a potential that grows with the dimensionless separation distance $S$. The global minimum at $S\approx 0$ corresponds to a `collision' of the two wings, and indeed simulations show that the follower rear-ends the leader for some initializations. 

By computing and visualizing the associated flow fields, we also determine some important mechanisms at work as the follower oscillates within the wave-like flow left by the leader. When the follower is at a position such that its vertical flapping velocity is instantaneously in-phase with the vertical flow of the leader's wake, a low average thrust results. This synchronization of follower-wake motions leads to a constructive interaction that generates a stronger collective wake, but also a weak thrust on the follower as its vertical velocity relative to the ambient fluid is decreased. In contrast, if the follower is at a position such that its vertical motion is out-of-phase with the leader's wake flow, we observe a destructive interaction, a weaker collective wake, higher follower-flow relative velocities, and a maximum thrust. The equilibrium states are intermediate in these extremes and correspond to positions where the average thrust of the follower is largely unchanged relative to swimming in isolation. Importantly, the stability of observed schooling states is assured by a specific follower-wake phase relationship. At downstroke, for example, the follower sees a downward flow just upstream, so that if it were perturbed forward, its thrust would decrease, countering the perturbation. Similarly, upward flows just downstream of the follower ensure that perturbations that increase separation yield increased thrust and again a return to the stable equilibrium position.

These explorations also suggest that, through appropriate positioning within a leader's wake, a follower may take advantage of flow interactions. We demonstrate this for the case of a weakly-flapping follower who may keep up with a faster-flapping leader by exploiting its wake flow. This ability of a pair of uncoordinated flapping wings to move together cohesively due to the interaction of the follower with the wake left by the leader was also observed experimentally by \citet{newbolt2019flow}. We find that this `freeloading' follower who flaps with smaller amplitude can travel up to four times faster than it would in isolation by taking up favorable positions in which it oscillates counter to the leader's wake flows. Further, the Froude efficiency of the follower is more than doubled and in some cases the follower's net energy is positive, which implies that energy could in principle be stored and used for other purposes. The leader is minimally affected, and so the efficiency of the pair is also improved.

For all these two-body interaction settings, the leader is affected much less, though some systematic trends are observed. We find that the leader's swimming speed, which is also that of the pair for the schooling states, decreases in comparison to isolated swimming. Interestingly, this decrease reflects an upstream influence due to the presence of the follower. When the follower is very close to the leader, the speed drops by as much as $50\%$, and this decrease is only $10\% \sim 20\%$ if the follower is one wavelength behind. The pair speed approaches the single wing speed for large separations. This result is opposite to that observed in recent experiments \citep{Ramananarivo2016Flowinteractionslead} and simulations~\citep{heydari2021school}, for which the leader's speed (and the pair speed) is somewhat faster than the single wing speed, and smaller separations are associated with higher speeds. Later investigations found that the flow interactions have negligible effects on the speed of the pair~\citep{newbolt2019flow,nitsche2024stability}.

Using a linearized theory, we have also calculated the hydrodynamic forces and the hydrodynamic potential on the follower wing. The linearized theory assumes a small amplitude of wing flapping and a small Strouhal number of the swimming motion, and calculates the interaction of the follower wing with the wake shed by the leader. This linearized theory shows the hydrodynamic force on the follower is a sinusoidal function of period one in the schooling number, which agrees with our vortex sheet simulations. Moreover, the maximum (minimum) of the hydrodynamic thrust produced by the follower is achieved when the wing-flow relative velocity achieves a maximum (minimum). An analytic form of the hydrodynamic potential is also calculated, from which the stable schooling numbers are solved and found to agree with the simulations for small Strouhal numbers. Both the simulations and the linearized theory agree well with the stable schooling numbers observed in experiments \citep{Ramananarivo2016Flowinteractionslead}. For two wings having different flapping amplitudes, the linearized theory produces a condition for the follower to keep up with the leader, exploiting its wake for faster propulsion.

The case of two flapping wings is perhaps the simplest `school' or `flock' of interacting locomotors, and many elaborations might be considered for future study. One direction is to consider the influence of differing flapping kinematics on the emergent configurations and dynamics of a group, whether a pair or larger. This current work has shown the re-positioning that results from changing flapping amplitude alone. The experiments in \citet{newbolt2019flow} showed that differing temporal phases lead to schooling states of different separations and different schooling numbers, and that stable positioning from differing flapping frequencies among individuals can still be achieved. Therefore, it would be interesting to derive a `keep-up' condition for this scenario as well. 
The two-wing system can also inform the understanding of more complex, many-body configurations and the behavior of more sophisticated locomotion systems.
In a tandem or single-file arrangement, a third wing interacts with a flow that is the superposition of flows by both upstream neighbors, for example. 
The extent to which interactions are pairwise versus cumulative was investigated by \citet{newbolt2024flow}, who found that pairwise flow interactions promote orderly arrays or `crystals' of individuals, but that this order can get disrupted by unstably growing positional waves. Results obtained
either experimentally or numerically for few interacting bodies~\citep{lin2020self,lin2024various,heydari2024mapping,nitsche2024stability,newbolt2024flow} 
 could still benefit from a rigorous analysis of the pairwise interactions.

Recent works have also explored pairs of self-propelled, inline plates with pitching motions. A numerical study by \citet{heydari2021school} revealed that these motions significantly enhance the group's swimming efficiency compared to when the plates swim in isolation. Such improvements in efficiency can also occur with mismatched foil pitching amplitudes~\citep{han2024tailoring}. A range of flow conditions, including phase synchrony and pitching amplitude, was shown to lead to two-dimensional stability of various formations in recent experiments on the self-organization of a pair of pitching hydrofoils~\citep{ormonde2024two}. 
Therefore, it would be interesting to explore two-dimensional configurations in which heaving-and-pitching individuals are spaced laterally and perhaps allowed to move laterally or reorient due to the flow interactions. These elaborations could shed light on classical models of 2D fish schools and indeed on the flows and dynamics of collectively swimming and flying animals.\\

\noindent \textbf{Acknowledgements}

We thank Sophie Ramananarivo, Anand Oza, Joel W.\ Newbolt, Steve Childress and Jun Zhang for fruitful discussions of the modeling, and thank Michael O'Neil, Sunli Tang and Abtin Rahimian for discussions of numerical quadratures. This work was supported by the Lyttle Chair in Applied Mathematics to M.S.\ and the U.S. National Science Foundation award DMS-1847955 to L.R. \\

\noindent \textbf{Declaration of interests}

The authors report no conflict of interest.

\appendix
\section{Vortex sheet model}\label{appsubsec: vortex sheet model}\label{sec:appendix}
In the vortex sheet simulation, high order quadrature rules for the Cauchy singular integrals (Eq.~(\ref{eq:w_kst})) must be applied to prevent the free vortex sheet of the leader wing penetrating the bound vortex sheet of the follower wing. In this section, we propose two quadrature rules for the bound sheet and free sheet, respectively. First, a singularity subtraction technique is proposed for the bound vortex sheet, which is accurate for polynomial approximations of the vortex sheet strength $\bar{\gamma}(s)$. This quadrature allows us to calculate the velocity induced by the bound vortex sheet (approximated by polynomials) accurately at any point off the boundary. Second, for the free vortex sheet, we analytically calculate the velocity induced by each vortex sheet segment, the sum of which provides the velocity field induced by the full free vortex sheet.

\subsection{Singularity subtraction method for the bound vortex sheet integral}
We first consider a general Cauchy integral of the form:
\begin{equation}
G(z)=\int_{-1}^{1}\frac{\gamma(s)}{z-s}\,\d s,\quad \text{for } z\notin[-1,1],\label{eq:int}
\end{equation}
where $z=x+iy$ is a complex point not on the finite segment $[-1,1],$ and $\gamma(s)$ is a real function. Eq.~(\ref{eq:int}) is also called the finite Hilbert transform of the real function $\gamma(s)$ \citep{King2009Hilberttransformsa,Olver2011ComputingHilberttransform}. In our vortex sheet model, if we assume that the position of the rigid bound vortex sheet is 
\begin{equation}
\xi(s)=c+\hat{s}s,\ \text{where }c=\xi(0),\ \text{ and }-1\le s\le1,
\end{equation}
then the velocity field induced by the bound vortex sheet (Eq.~(\ref{eq:w_kst})) is expressed as
\begin{equation}
I_{b}(z)=\frac{1}{2\pi i}\int_{-1}^{1}\frac{\gamma(s)}{z-\xi(s)}\,\d s ={\frac{1}{2\pi i}} \frac{1}{\hat{s}}\int_{-1}^{1}\frac{\gamma(s)}{(z-c)/\hat{s}-s}\,\d s={\frac{1}{2\pi i}}\frac{1}{\hat{s}}G\left(\frac{z-c}{\hat{s}}\right).
\end{equation}
Here $\hat{s}$ denotes the tangential direction of the wing. In the following subsection, we show the singularity subtraction method for calculating the Hilbert transform  Eq.~(\ref{eq:int}).

\subsubsection{$\gamma(s)$ is a polynomial}
Assume $\gamma(s)$ is a polynomial, which could be a polynomial interpolation of some function. In Eq.~(\ref{eq:int}), we separate the singular part manually by subtracting the constant $\gamma(z)$ from $\gamma(s).$ The
integral then splits into two parts,
\begin{eqnarray*}
G(z)=\int_{-1}^{1}\frac{\gamma(s)}{z-s}\,\d s & = & -\int_{-1}^{1}\frac{\gamma(s)-\gamma(z)}{s-z}\,\d s+\gamma(z)\int_{-1}^{1}\frac{1}{z-s}\,\d s\\
 & = & -\int_{-1}^{1}g(s)\,\d s-\gamma(z)\log\frac{1-z}{1+z}.
\end{eqnarray*}
The first part is a regular integral of the smooth function $g(s)$, given by
\begin{equation}
g(s)=\frac{\gamma(s)-\gamma(z)}{s-z}.
\end{equation}
In fact, if we assume $\gamma(s)$ is a polynomial of a real variable $s$ with real coefficients, then $g(s)$ is a polynomial of real $s$, while the coefficients are  polynomials of complex variable $z$. Therefore, high order quadrature rules such as Gauss-Legendre quadrature or Clenshaw-Curtis quadrature can be applied, both of which provide accurate evaluations for polynomials \citep{Trefethen2008IsGaussquadrature}. 

\subsubsection{$\nu(s)=\gamma(s)\sqrt{1-s^{2}}$ is a polynomial}
In the vortex sheet simulations, the true vortex sheet strength contains an inverse-square-root singularity at the wing leading edge, which
can be expressed as
\begin{equation}
\gamma(s)=\frac{\nu(s)}{\sqrt{1-s^{2}}},\quad -1\le s\le1,
\end{equation}
where $\nu(s)$ can be a smooth function. Now we consider the following singular integral
\begin{equation}
G(z)=\int_{-1}^{1}\frac{\nu(s)}{\sqrt{1-s^{2}}\left(z-s\right)}\,\d s,\quad \text{for } z\notin[-1,1],
\end{equation}
and assume $\nu(s)$ is a polynomial. In our vortex sheet method, it is a polynomial interpolation on Chebyshev-Gauss-Lobatto nodes (Eq.~(\ref{Eq:Lobatto})).

Using the singularity subtraction trick, we have
\begin{eqnarray*}
G(z) = \int_{-1}^{1}\frac{\nu(s)}{\sqrt{1-s^{2}}\left(z-s\right)}\,\d s
 & = & -\int_{-1}^{1}\frac{1}{\sqrt{1-s^{2}}}\frac{\nu(s)-\nu(z)}{s-z}\,\d s+\int_{-1}^{1}\frac{\nu(z)}{\sqrt{1-s^{2}}\left(z-s\right)}\,\d s\\
 & = & -\int_{-1}^{1}\frac{g(s)}{\sqrt{1-s^{2}}}\,\d s+\nu(z)\int_{-1}^{1}\frac{1}{\sqrt{1-s^{2}}\left(z-s\right)}\,\d s,
\end{eqnarray*}
where $g(s)=({\nu(s)-\nu(z)})/({s-z})$ is a polynomial of $s$ with complex coefficients that are polynomials of $z$. We evaluate the integral of $g(s)$ using Chebyshev quadrature at Chebyshev-Gauss-Lobatto nodes (Eq.~(\ref{Eq:Lobatto})), which is accurate for polynomials \citep{Mason2002Chebyshevpolynomials}. The second term can be calculated analytically \citep[][p.~201]{Golberg2013Numericalsolutionintegral}:
\begin{equation}
\int_{-1}^{1}\frac{1}{\sqrt{1-s^{2}}\left(z-s\right)}\,\d s = \frac{\pi}{\sqrt{z+1}\sqrt{z-1}},\quad \text{for } z\notin[-1,1].
\end{equation}

\subsection{Quadrature on the free vortex sheet}
Now we discuss the quadrature on the free vortex sheet $C_{f}$: $\ s_{\min}\le s\le s_{\max}$, for evaluating the potential flow induced by this sheet. Denote the position of the sheet by $\xi(s)$ and the arc-length by $s$. Consider the following integral on $C_f$ for the smooth real function $\gamma(s)$:
\begin{equation}\label{eq:Ifree}
I(z)=\int_{s_{\min}}^{s_{\max}}\frac{\gamma(s)}{z-\xi(s)}\,\d s,\quad z\notin[-1,1].
\end{equation}
We discretize the boundary $C_f$ at $\xi_{k}=\xi(s_{k}),\ k=1,\ldots,N$, and let
\begin{equation}
\gamma_k = \int_{s_{k}}^{s_{k+1}}\gamma(s)\,\d s/(s_{k+1}-s_k)
\end{equation} denote the average of $\gamma$ on each piece.
The boundary $C_f$ can be approximated by a piecewise linear segment,
\begin{equation}
\tilde{\xi}(s)\simeq p_{k}s+q_{k},\;\; \text{for }s_{k}\le s\le s_{k+1},
\end{equation}
where $\tilde{\xi}(s_k)=\xi_k$ are the endpoints of the each segment. On each segment, $\gamma(s)$ is approximated by a piecewise constant function, 
\begin{equation}
\tilde{\gamma}(s)\simeq\gamma_{k},\;\; \text{for }s_{k}\le s\le s_{k+1}.
\end{equation}
Next, we approximate the integral Eq.~(\ref{eq:Ifree}) by the integral of $\tilde{\gamma}$ on $\tilde{\xi}$, which can be evaluated analytically. The integral Eq.~(\ref{eq:Ifree}) can be written as the sum of integral contributions on each discrete piece, and then be approximated, as
\begin{equation}\label{eq:Ikfree}
I(z)=\int_{s_{\min}}^{s_{\max}}\frac{\gamma(s)}{z-\xi(s)}\,\d s=\sum_{k=1}^{N-1}\int_{s_{k}}^{s_{k+1}}\frac{\gamma(s)}{z-\xi(s)}\,\d s=\sum_{k=1}^{N-1}I_k(z),
\end{equation}
where 
\begin{equation}
I_{k}(z)=\int_{s_{k}}^{s_{k+1}}\frac{\gamma(s)}{z-\xi(s)}\,\d s\simeq\int_{s_{k}}^{s_{k+1}}\frac{\gamma_{k}}{z-q_{k}-p_{k}s}\,\d s=\frac{\gamma_{k}}{p_{k}}\log\left(\frac{z-\xi_{k}}{z-\xi_{k+1}}\right),\label{eq:Ik}
\end{equation}
and $p_{k}=(\xi_{k+1}-\xi_{k})/(s_{k+1}-s_{k})$. Therefore, the integral on the boundary Eq.~(\ref{eq:Ifree}) is approximated as
\begin{equation}
I(z)\simeq\sum_{k=1}^{N-1}\frac{(s_{k+1}-s_{k})\gamma_{k}}{\xi_{k+1}-\xi_{k}}\log\left(\frac{z-\xi_{k}}{z-\xi_{k+1}}\right).
\end{equation}
We note that $(s_{k+1}-s_{k})\gamma_{k}$ denotes the circulation on the vortex sheet segment $[\xi_k,\xi_{k+1}]$, which is the unnormalized vortex sheet strength; see Section~\ref{sec: simulation method}.

\bibliographystyle{jfm}
\bibliography{schooling_paper_jfm}

\end{document}